\documentclass[floatfix,superscriptaddress,a4paper,
               nofootinbib,preprint]{revtex4-1}
\usepackage[colorlinks=true,linktocpage=true,linkcolor=blue,citecolor=blue,allcolors=blue]{hyperref}
\pdfoutput=1

\usepackage{graphicx}
\usepackage{amsmath}
\usepackage{amssymb}
\usepackage{bm}

\newcommand{\hsp}{\hspace*{1pt}}
\newcommand{\hspm}{\hspace*{.5pt}}
\newcommand{\ds}{\displaystyle}
\newcommand{\be}{\begin{equation}}
\newcommand{\ee}{\end{equation}}
\newcommand{\bel}[1]{\be\label{#1}}
\newcommand{\re}[1]{Eq.~(\ref{#1})}

\usepackage{color}

\begin{document}

\title
{
Phase diagram of alpha matter\\
with Skyrme-like scalar interaction
}

\author{L.~M. Satarov}
\affiliation{
Frankfurt Institute for Advanced Studies, D-60438 Frankfurt am Main, Germany}

\author{R. V. Poberezhnyuk}
\affiliation{
Frankfurt Institute for Advanced Studies, D-60438 Frankfurt am Main, Germany}
\affiliation{
Bogolyubov Institute for Theoretical Physics, 03680 Kiev, Ukraine}

\author{I.~N. Mishustin}
\affiliation{
Frankfurt Institute for Advanced Studies, D-60438 Frankfurt am Main, Germany}
\affiliation{
National Research Center ''Kurchatov Institute'' 123182 Moscow, Russia}

\author{H. Stoecker}
\affiliation{
Frankfurt Institute for Advanced Studies, D-60438 Frankfurt am Main, Germany}
\affiliation{
Institut f\"ur Theoretische Physik,
Goethe Universit\"at Frankfurt, D-60438 Frankfurt am Main, Germany}
\affiliation{
GSI Helmholtzzentrum f\"ur Schwerionenforschung GmbH, D-64291 Darmstadt, Germany}

\begin{abstract}
The equation of state and phase diagram of strongly interacting matter composed
of $\alpha$~particles are studied in the mean-field approximation.
The particle interactions are included via a Skyrme-like mean field, containing both
attractive and repulsive terms. The
model parameters are found by fitting the values of binding energy and
baryon density in the ground state
of~$\alpha$ matter, obtained from microscopic calculations by Clark and Wang~[\href{https://www.sciencedirect.com/science/article/abs/pii/0003491666902363?via\%3Dihub}{Ann.~Phys.~(NY)~{\bf 40}, 127 (1966)}]. Thermodynamic quantities
of $\alpha$~matter are calculated in the broad domains of temperature and baryon density, which can be
reached in heavy-ion collisions at intermediate energies. The model predicts both first-order liquid-gas
phase transition and Bose-Einstein
condensation of~$\alpha$ particles. We
present the profiles of scaled variance,
sound velocity and isochoric heat capacity along the isentropic trajectories of~$\alpha$ matter.
Strong density fluctuations are predicted
in the vicinity of the critical point
at temperature $T_c\approx 14~\textrm{MeV}$
and density $n_c\approx 0.012~\textrm{fm}^{-3}$.
\end{abstract}

\maketitle

\section{Introduction}

It is well known from experimental observations \cite{Schuttauf:1996ci,Reisdorf:2010aa,Wada:2019utj} and
theoretical studies \cite{Gro83,Bon85,Peilert:1991sm,Bondorf:1995ua} that symmetric nuclear matter at
subsaturation densities has a tendency for clusterization. And $\alpha$ particles are the most abundant clusters,
as observed in many experiments \cite{Marini:2015zwa,Borderie:2016tmc}.
The fact that $\alpha$ particles are bosons makes $\alpha$ matter even more interesting to study, because of
the possibility of Bose-Einstein condensation (BEC).

A special role of $\alpha$ particles is well established in nuclear physics.
The examples include: $2\hspm\alpha$ structure of $^8$Be  which decays into two
$\alpha$'s in about $10^{-16} s$ and the famous $3\hspm\alpha$ Hoyle state in $^{12}$C\, with width of only $8.5$~eV.
Recently, \mbox{$\alpha$-clustered} excited states, analogous to the Hoyle state,
have been identified in $^{16}$O and $^{24}$Mg~\cite{Sch17,Art20}. In  Ref.~\cite{Bar18}
an evidence for $7\hspm\alpha$ resonance has been reported.
All these states can be interpreted as Bose condensates of $\alpha$ particles in finite systems~\cite{Sch16,Cao20}.

Early theoretical studies of clusterized nuclear matter applied
variational calculations with phenomenological two-body potentials. For example,
in Ref.~~\cite{Cla66} Clark and Wang calculated energy per baryon for
pure $\alpha$ matter by using
several potentials of $\alpha\alpha$ interaction. They have found that such matter has
a ground state (GS) with the binding energy, only several MeV (per baryon) smaller than the iso-symmetric nuclear matter. It was demonstrated
~\cite{Joh80} that this binding energy is sensitive to the short-range behavior of the potential.

One should also mention the
virial~\cite{Hor06} and quasiparticle~\cite{Sed17,Sed20} models. In particular, in
Ref.~\cite{Hor06}, the authors  express the pressure and partial densities of \mbox{$\alpha-N$} matter via integrals over
observed phase shifts of $NN,  \alpha N$ and $\alpha\alpha$~scat\-tering. Although such an approach is 'model-independent',
it can be applied only for dilute matter.

Significant progress in modern computing has stimulated 'ab-initio'
microscopic calculations of nuclei, which use realistic $2N$ and $3N$ forces to describe
cluster degrees of freedom
(for a recent review, see Ref.~\cite{Fre18}). Such studies include, e.g., the
antisymmetrized~\cite{Hor91} and fermionic~\cite{Fel90} molecular dyna\-mics,
the quantum Monte Carlo approach~\cite{Pie92} and the nuclear lattice effective \mbox{theory~\cite{Epe11,Elh16}}.
In particular,  it has been shown in Ref.~\cite{Elh16} that increasing the depth of the nucleon-nucleon potential at short distances
(without significantly changing the $NN$ and $3N$ data) leads to formation of a clusterized phase in GS of $\alpha$-congugate 
nuclei ($N=Z\geqslant 4$). The above-mentioned microscopic methods are computationally demanding and currently may deal only with light- and medium-size nuclei. On the other hand, they consistently take into account short-range correlations of nucleons
and in-medium modification of clusters.

Recently another microscopic method for studying clusterization phomena in nuclei has been
proposed~\cite{Ebr12,Ebr14,Ebr20}, namely the nuclear density functional \mbox{theory~(DFT)}.
In fact, it is an~extension of the mean-field approach.
Within the DFT, clusters are associated with spatially localized fluctuations of nucleon density.
As demonstrated in Ref.~\cite{Ebr14}, main features of $\alpha$-conjugate
nuclei can be well described within this approach.

Several versions of the relativistic mean-field (RMF) model have been proposed to describe
clusterized nuclear matter (see, e.g., Refs.~\cite{Mis16,Pai18,Zha19}). Due to neglect of NN correlations, clusters are
introduced in this model as separate degrees of freedom. This leads to appearance of additional model parameters
characterizing interactions of clusters and nucleons.
In Refs.~\cite{Pai18,Zha19}, the authors additionally introduce shifts of cluster masses to implement disappearance of clusters at high baryon densities.

The possibility of the Bose-Einstein cluster condensation in stellar matter at nonzero temperatures
has been studied within the
quantum-statistical model in Ref.~\cite{Fur20}.  Attempts to implement quantum statistics
for nucleons and nuclear clusters have been also made
in the quasiclassical molecular dynamical model~\mbox{\cite{Peilert:1991sm,Pei92,Mar98}}. This was
achieved by adding the phenomenological
Pauli potential which becomes strongly repulsive
for nucleon pairs with close phase-space coordinates.
In Ref.~\cite{Mar98} a resonable desciption of light cluster properties
has been obtained within this approach.

In our previuos works~\cite{Sat17,Sat19,Sat20}  several mean-field models
with Skyrme-like vector potentials have been developed to describe simultaneously
the liquid-gas phase transition  (LGPT) and BEC in $\alpha$ and $\alpha-N$ matter.
In the present paper, we propose another  approach where
$\alpha$-clustered nuclear matter is described by a scalar field $\phi$  with an effective
Lagrangian containing the attractive~($\phi^4$) and repulsive~($\phi^6$) terms.
Analogous scalar potential was first introduced by Boguta \textit{et al.} in the RMF model
of nuclear matter~\cite{Bog77,Bog83}.  In Ref.~\cite{Mis19}, a similar model was used for bosonic particles
without any conserved charge, i.e., with zero chemical potential (e.g., pions).
In the present paper we extend this approach for $\alpha$ particles which carry the
conserved baryon charge.

The paper is organized as follows. In Section~II we formulate
the model. In Sec.~III we
analyze the phase diagram of $\alpha$ matter which contains both LGPT and BEC.
We also discuss isentropic and isothermal trajectories of $\alpha$ matter in different
thermodynamic variables.
In~Sec.~IV A we present a~fluctuation observable in terms of the scaled variance~$\omega$, and show
that it has a~strong peak at the critical point of liquid-gas phase transition.
\mbox{In ~Sec.~IV B} and IV C we calculate isentropic profiles of the sound velocity
and the isochoric heat capacity. It is shown that
both these quantities exhibit jumps at the mixed phase boundary.

\section{The model}
\label{sec-model}

\subsection{Motivation}

In Refs.~\cite{Sat19,Sat20}, we have studied nuclear systems composed of $\alpha$ particles and nucleons.
Two possibilities regarding the interaction between $\alpha$
particles and nucleons have been considered.  In the first scenario~\cite{Sat19}, pure $\alpha$ matter
and pure nucleonic matter have their own GSs separated by a potential
barrier. In this case the resulting phase diagram has two LGPTs: one (metastable) in the domain with large $\alpha$
abundances and another one (stable) in the domain dominated by nucleons. In Ref.~\cite{Sat20} we have 
analyzed the sensitivity of nuclear matter phase diagram to the strength of $\alpha N$  interaction.
Within the mean-field approximation (MFA), a threshold-like behavior of $\alpha$ clustering
in the cold nuclear matter has been predicted. This effect is similar to the 'quantum phase 
transition' discovered in Ref.~\cite{Elh16}. We show that the energy barrier separating 
the clusterized- and non-clusterized phases disappears at some critical strength of the $\alpha N$ attraction.          
In this case, there is only one common GS with coexis\-ting nucleons and $\alpha$'s in chemical equilibrium, 
but the admixture of $\alpha$~particles is small at
moderate temperatures and densities.

Having in mind that experimental data on heavy-ion collisions at intermediate
energies~\cite{Sch17,Bar18}
show enhanced yields of $\alpha$~particles and $\alpha$-conjugate nuclei, we
are tempted to conclude that the first scenario is more realistic.
In this case, there is a~chance that the excited matter produced in such collisions will
expand along  a~metastable branch of the phase diagram where $\alpha$ particles
are more abundant and even may form a~condensate.  In the present paper, we study this
scenario  within a~Skyrme-type mean-field model for pure $\alpha$ matter,
i.e., ignoring the admixture of nucleons.

\subsection{Bosonic matter in the mean-field approximation}
\label{sec-bosm}

Following~Ref.~\cite{Mis19} we describe the system of scalar bosons
with zero spin\hspm\footnote
{
All numerical calculations below are carried out
for the specific case of $\alpha$ particles.
}
by a scalar field operator $\phi\hspm (x)$ with the Lagrangian ($\hbar=c=1$)
\bel{lag1}
{\cal L}=\frac{1}{2}\left(\partial_\mu\phi\,\partial^{\,\mu}\phi -m^2\hspm\phi^2\right)+{\cal L}_{\rm int}\left(\phi^2\right).
\ee
Here $m$ is the boson mass in the vacuum and $\cal{L}_{\rm int}$ contain particles' interactions.
In the MFA one expands $\cal{L}_{\rm int}$ in the lowest order in $\phi^2-\sigma$,
where
\bel{sig1}
\sigma=\langle\phi^2\rangle
\ee
is the scalar density ($c$-number) and angular brackets denote the quantum-statistical ave\-raging in
the grand-canonical ensemble (see below).

One gets the following result
\bel{lag2}
{\cal L}\approx{\cal L}_{\rm MFA}=\frac{1}{2}\left[\partial_\mu\,\phi\,\partial^{\,\mu}\phi -M^2(\sigma)\hsp\phi^2\right]
+p_{\hsp\rm ex}(\sigma)\,,
\ee
where the quantity
\bel{efm1}
M^2(\sigma)=m^2- 2{\cal L}^\prime_{\rm int}(\sigma)
\ee
is the effective mass squared, and
\bel{prex}
p_{\hsp\rm ex}(\sigma)={\cal L}_{\rm int}(\sigma)-\sigma {\cal L}^\prime_{\rm int}(\sigma)
\ee
is the so-called excess pressure~\cite{Mis19} (hereinafter primes denote derivatives over $\sigma$)\hspm.
The~terms containing ${\cal L}_{\rm int}$ in these two equations describe deviations from
the ideal gas. Later on we apply the Skyrme-like parametrization for
${\cal L}_{\rm int}(\sigma)$ containing both the attractive and repulsive parts.
From Eqs.~(\ref{efm1}), (\ref{prex}) one gets the differential relation
between the effective mass and excess pressure
\bel{prex1}
p_{\hsp\rm ex}^{\,\prime}(\sigma)=\sigma M M^\prime(\sigma)\,,
\ee
which   {guaranties} the thermodynamic consistency
of the present model   {(see below)}. In fact, the appearance of the effective mass makes
the main difference between the present approach and our previous studies
of $\alpha$~\cite{Sat17} and $\alpha-N$~\cite{Sat19,Sat20} matter.

Using a plane-wave decomposition of $\phi$ in terms of creation ($a_{\bm{k}}^+$) and
annihilation ($a_{\bm{k}}$) operators\hspm\footnote
{
For states with BEC one should add~\cite{Mis19} an additional condensate component $\phi_c$
to the right-hand side (r.h.s.) of this equation.
}
\bel{fdec}
\phi\hspm (x)=\frac{g}{(2\pi)^3}\int \frac{d^{\,3} k}{\sqrt{2E_k}}\left(a_{\bm{k}} e^{-ikx}+a_{\bm{k}}^+ e^{ikx}\right)\,,
\ee
where $k^0=E_{\bm{k}}=\sqrt{M^2+{\bm{k}}^2}$ and $g$ is the statistical weight of a boson particle, one gets~\cite{Mis19} the
following equations for the   {particle} number and Hamiltonian density operators in the~MFA
\bel{hamd}
\frac{N}{V}=\frac{g}{(2\pi)^3}\int d^{\,3}k\hsp a_{\bm{k}}^+a_{\bm{k}}\,,~~~
\frac{H}{V}=\frac{g}{(2\pi)^3}\int d^{\,3}k\hsp E_{\bm{k}}\hsp a_{\bm{k}}^+a_{\bm{k}} -p_{\rm ex}\,.
\ee

In the grand canonical ensemble one can find the particle momentum distribution by averaging $a_{\bm{k}}^+a_{\bm{k}}$ over
the statistical operator \mbox{$\rho\propto\exp\left[(\mu N-H)/T\right]$} at given chemical potential~$\mu$ and temperature $T$. Within the MFA one has
\bel{pmd1}
n_{\bm{k}}=\langle a_{\bm{k}}^+a_{\bm{k}}\rangle=\left[\exp\left(\frac{E_{\bm{k}}-\mu}{T}\right)-1\right]^{-1}\,.
\ee
One can see that this distribution coincides with the ideal-gas (Bose-Einstein) distribution of   {quasi}particles with mass $M$\hspm . The latter
is in general not equal to $m$ and should be found self-consistently at given $T$ and $\mu$. According to \re{pmd1} possible values of $\mu$
are bound from above, namely $\mu\leqslant M$. In the particular case $\mu=M$ the distribution $n_{\bm{k}}$ is singular at~\mbox{$\bm{k}=0$}\,.
Similarly to the ideal-gas case, this implies the appearance of BEC with macroscopic number of zero momentum particles.
Therefore, in our model the condensate appears at
\mbox{$\mu=M(\sigma)$}. We will see that this takes place either at low enough temperatures or at sufficiently large densities.

Using Eqs.~(\ref{sig1}), (\ref{fdec}), (\ref{pmd1}) one can calculate the scalar density $\sigma$:
\bel{gape}
\sigma=\sigma_{\rm th}\left[T,\mu,M(\sigma)\right],
\ee
where
\bel{gape1}
\sigma_{\rm th}\left(T,\mu,M\right)=\frac{g}{(2\pi)^3}\int d^{\,3}k\hsp\frac{n_{\bm{k}}}{E_{\bm{k}}}=
\frac{g}{2\pi^2}\int\limits_M^\infty\frac{dE\sqrt{E^2-M^2}}{\exp{\left(\frac{\ds E-\mu}{\ds T}\right)-1}}\,.
\ee
In fact, \re{gape} plays a role of self-consistent gap equation which determines $\sigma$ and $M$ as functions of~$T,\mu$.

The pressure $p$ is determined by spatial components of the energy-momentum tensor $T_{\alpha\alpha}$ which
in turn can be calculated from the Lagrangian. Within the MFA   {one has}
\bel{pre1}
p=\frac{1}{3}\left<T_{xx}+T_{yy}+T_{zz}\right>=\frac{1}{6}\left<\left(\bm
\nabla\phi\right)^2\right>+p_{\,\rm ex}(\sigma)\,.
\ee
Using Eqs.~(\ref{fdec}), (\ref{pmd1}) one can easily calculate the kinetic term in the second equality (below we denote it
by $p_{\,\rm th}$). Finally one gets the equation
\bel{pre2}
p=p_{\hsp\rm th}\left(T,\mu,M\right)+p_{\hsp\rm ex}(\sigma)\,,
\ee
where
\bel{pre3}
p_{\hsp\rm th}=\frac{g}{(2\pi)^3}\int d^{\,3}k\,\frac{\bm k^2}{3E_{\bm{k}}}\hsp n_{\bm{k}}=
\frac{g}{6\pi^2}\int\limits_M^\infty\frac{dE\hspm \left(E^2-M^2\right)^{3/2}}{\exp{\left(\frac{\ds E-\mu}{\ds T}\right)}-1}\,.
\ee

One can also obtain explicit expressions for the number density of $\alpha$ particles
\mbox{$n=\langle N \rangle\hspace*{-2pt}/V$}
and the internal energy density
\mbox{$\varepsilon=\langle H \rangle\hspace*{-2pt}/V$}  (note that the baryon density $n_B=4\hspm n$\hspm). Using Eqs.~(\ref{hamd}), (\ref{pmd1})
one has
\bel{vden}
n=n_{\hsp\rm th}\left(T,\mu,M\right)=\frac{g}{(2\pi)^3}\int d^{\,3}k\, n_{\bm{k}}\,,
\ee
and
\bel{eden}
\varepsilon=\varepsilon_{\hsp\rm th}\left(T,\mu,M\right)-p_{\hsp\rm ex}(\sigma),
~~~~~\varepsilon_{\hsp\rm th}=\frac{g}{(2\pi)^3}\int d^{\,3}k\hsp E_{\bm{k}}
\hsp n_{\bm{k}}\,.
\ee

In the case of $\alpha$ matter, the temperatures of interest are lower or of  the
order of \mbox{$\alpha$-particle} binding energy ($\approx$ 28 MeV).
To a good accuracy, one can apply the non-relativistic
approximation (NRA) by taking the lowest-order terms in $T/M$ in Eqs.~(\ref{gape1}), (\ref{pre3})--(\ref{eden}).
This leads to approximate relations
\begin{eqnarray}
&&\sigma_{\rm th}\approx\frac{n_{\,\rm th}}{M}\approx\frac{g}{M\lambda_T^3}\,g_{\,3/2}\hspm (z)\,,\label{snnr}\\
&&p_{\hsp\rm th}\approx\frac{2}{3}\left(\varepsilon_{\,\rm th}-M\hspm n_{\,\rm th}\right)
\approx\frac{g\hspm T}{\lambda_T^3}\,g_{\,5/2}\hspm (z)\,,\label{penr}
\end{eqnarray}
where $z\leqslant 1$ is a nonrelativistic fugacity, and $\lambda_T=\lambda_T\hsp (T,M)$ is the thermal wave length:
\bel{sntr1}
z\equiv\exp\left(\frac{\mu-M}{T}\right),~~~~\lambda_T
 =
\sqrt{\frac{2\pi}{M\hsp T}}\,\,.
\ee
The dimensionless function (polylogarithm) $g_{\hspm\beta}\hspm (z)$ is
defined as
\bel{bint}
g_{\hspm\beta}\hspm (z)\equiv\frac{1}{\Gamma(\beta)}\int\limits_0^\infty dx\hsp\frac{x^{\beta-1}}{z^{-1}e^{\ds x}-1}
=\sum_{k=1}^{\infty}z^k k^{-\beta}\,,
\ee
where $\Gamma(\beta)$ is the gamma function.

The following properties of the polylogarithm will be used below:
\bel{plpro}
z\hspm g_{\hspm\beta}^{\,\prime}\hspm (z)=g_{\hspm\beta-1}\hspm (z),~~~
g_{\hspm\beta}\hspm (z)=\left\{
\begin{array}{ll}z,&z\ll 1\hsp ,\\
\xi (\beta),&~z\to 1\hsp ,
\end{array}\right.
\ee
where $\xi (\beta)=\sum_{k=1}^{\infty}k^{-\beta}$ is the Riemann function\,.
The function $g_{\hspm\beta}\hspm (z)$ diverges at $z\to 1$
if $\beta\leqslant 1$\,. The classical (Boltzmann) limit
corresponds to small $z$. According to Eqs.~(\ref{snnr}) and~(\ref{plpro}),
in this case $z\approx n\lambda_T^3/g\ll 1$ and $p_{\,\rm th}\approx nT$.

On the other hand, for states where the degeneracy parameter
$n\lambda_T^3\gtrsim 1$, the effects of quantum statistics are important.
As discussed above, the BEC starts when $\mu\to M$\hspm   {, i.e., at}
$z\to 1$\,. From
Eqs.~(\ref{snnr}),~(\ref{sntr1}) and (\ref{plpro}) one can see that this
happens at~$T< T_{\rm BEC}$ where
\bel{tbec}
T_{\rm BEC}\approx\frac{2\pi}{\mu}\left(\frac{n}{g\hsp\xi\hspm (3/2)}\right)^{2/3}
\ee
is the threshold temperature of BEC. The chemical potential in the r.h.s.
of this equation is determined from the relation $\mu=M\hspm (\sigma)$.
At not too high densities one can take $\mu\approx m$  with a good accuracy.
The resulting value of $T_{\rm BEC}$ coincides with that for the ideal
Bose gas.   Therefore, for nonrelativistic bosons
(like $\alpha$-particles) the BEC onset line in~the~$(n,T)$ plane is
the same as in the ideal gas
~\cite{Bay01,Sat17}.

At $T<T_{\rm BEC}$ additional condensate terms
should be added
 to the r.h.s. of Eqs.~(\ref{gape}), (\ref{vden}) and (\ref{eden})\hspm\footnote
{
  {By presence of BEC} the pressure is modified only indirectly, via the condensate scalar
density $\sigma_c$ in the
term~$p_{\,\rm ex}(\sigma)$\,.
}:
\bel{condt}
\sigma\hspm (\mu)=\sigma_c+\sigma_{\rm th}\hspm (T,\mu,\mu),
~~n=n_c+n_{\rm th}\hspm (T,\mu,\mu),~~\varepsilon=\varepsilon_c+\varepsilon_{\,\rm th}\hspm (T,\mu,\mu)-p_{\,\rm ex}(\sigma)\,,
\ee
where $n_c=\mu\hspm\sigma_c, \varepsilon_c=\mu\hspm n_c$ and $\sigma\hspm (\mu)$
is determined by solving the equation $\mu=M\hspm (\sigma)$\hspm . Note that $(\mu,T)$ dependence
of the condensate terms is fully determined by the first equality in~(\ref{condt})\,.

Let us consider now the entropy of a bosonic system $S$. The entropy density
$s=S/V$ can be calculated by using the general relation
$s=(\varepsilon+p-\mu\hspm n)/T$. One can see
that the excess- and (possible) condensate terms in $\varepsilon, p$ and $n$ are
cancelled and we arrive at the equation:
\bel{entd}
s=\frac{\varepsilon_{\,\rm th}+p_{\,\rm th}-\mu\hspm n_{\,\rm th}}{T}\,.
\ee
Formally, we get the same expression as for the ideal Bose gas~\cite{Mis19}.
However, the interaction effects enter via the effective mass $M$\,.
Substituting $\mu=M+T\hsp\ln{z}$ and using Eqs.~(\ref{snnr}),~(\ref{penr}) one obtains
the following expression for the entropy per particle $\widetilde{s}=s/n$
\bel{spent}
\widetilde{s}=\frac{S}{N}\approx\left\{\begin{array}{ll}
\dfrac{5}{2}\hsp\dfrac{g_{\,5/2}\hspm (z)}{g_{\,3/2}\hspm (z)}-\ln{z}\,,~~~&T>T_{\rm BEC}\,,\\[3mm]
\dfrac{5}{2}\hspm\dfrac{\xi\hspm (5/2)}{n\lambda_T^3(T,\mu)},&T<T_{\rm BEC}\hsp .
\end{array}\right.
\ee
Again, we arrive at the relations for the ideal Bose \mbox{gas~\cite{Pat11}},
but with the modified (state-dependent) particle mass. These results show
that the specific entropy $\widetilde{s}$
is constant at the BEC boundary
($z=1$):
\bel{sdbec}
\widetilde{s}\approx \dfrac{5}{2}\hspm\dfrac{\xi\hspm (5/2)}
{\xi\hspm (3/2)}\approx 1.284~~~~(T=T_{\rm BEC})\,.
\ee
This means that this boundary is an isentrope and the bosonic matter can not cross it during the isentropic evolution.
For example, it is not possible to reach the BEC region
by an adiabatic expansion of $\alpha$ matter from non-condensed initial states\hspm\footnote
{
However, this is possible if the bosonic matter contains
also the LGPT (see Sec.~\ref{sec-phase-diagr}).
}.
Note that the threshold value (\ref{sdbec}) does not
depend on the particle mass and the interaction parameters.

Within the MFA one obtains the standard thermodynamic relations~\cite{Lan75}
\begin{eqnarray}
&&dp=sdT+nd\mu\,,\label{dpre}\\
&&d\varepsilon=Tds+\mu dn\,.\label{deps}
\end{eqnarray}
To prove~(\ref{dpre}), we directly calculate the derivatives entering the pressure
differential
\mbox{$dp=dT\,\partial\hspm p/\partial T+d\mu\,\partial\hspm p/\partial\mu
+dM\,\partial\hspm p/\partial M$}. Using~Eqs.~(\ref{pre2})--(\ref{eden}), (\ref{entd})
one can check that $\partial p/\partial T=\partial\hspm p_{\,\rm th}/\partial T=s,
~\partial\hspm p/\partial\mu=\partial\hspm p_{\,\rm th}/\partial\mu=n_{\,\rm th}$
and~\cite{Mis19}
\bel{mder}
\frac{\partial\hspm p}{\partial M}=\frac{\partial\hsp (p_{\hsp\rm th}+p_{\hsp\rm ex})}{\partial M}
=-M\sigma_{\rm th}+M\sigma=\left\{\begin{array}{ll}
0\,,~~~~&T>T_{\rm BEC}\,,\\[3mm]
\mu\sigma_c,&T<T_{\rm BEC}\hsp ,
\end{array}\right.
\ee
where we have used Eqs.~(\ref{prex1}) and (\ref{condt}) (in the second and third equalities, respectively).
Finally we arrive at \re{dpre}.  Note that
in the BEC region we substitute $M=\mu$ and $\mu\sigma_c=n-n_{\,\rm th}$\,. Equation (\ref{deps}) can be obtained
from (\ref{dpre}) by differentiating the relation $\varepsilon=Ts+\mu\hsp n-p$\,.

\subsection{Skyrme-like parametrization of particle interactions in $\alpha$ matter}

The above results are obtained in the MFA,
and they do not depend on a specific form
of the interaction Lagrangian. The following calculations are carried out with
the Skyrme-like ($\phi^4-\phi^6$) parametrization,
\bel{skin}
\mathcal{L}_{\rm int}(\sigma)=\frac{a}{4}\hsp\sigma^2-\frac{b}{6}\hsp\sigma^3\,,
\ee
where $a,b$ are positive constants. The first and the \mbox{second} terms in (\ref{skin})
describe, respectively, the attractive and the repulsive interactions between the scalar
bosons. It will be shown below that the presence of attraction leads to the
appearance of the first-order (liquid-gas) phase transition of bosonic matter.
On the other hand, the repulsive term stabilizes this matter at   {high densities}\,.
%------------------------
\begin{figure}[htb!]
\centering
\includegraphics[trim=2cm 7.5cm 3cm 8cm,width=0.6\textwidth]{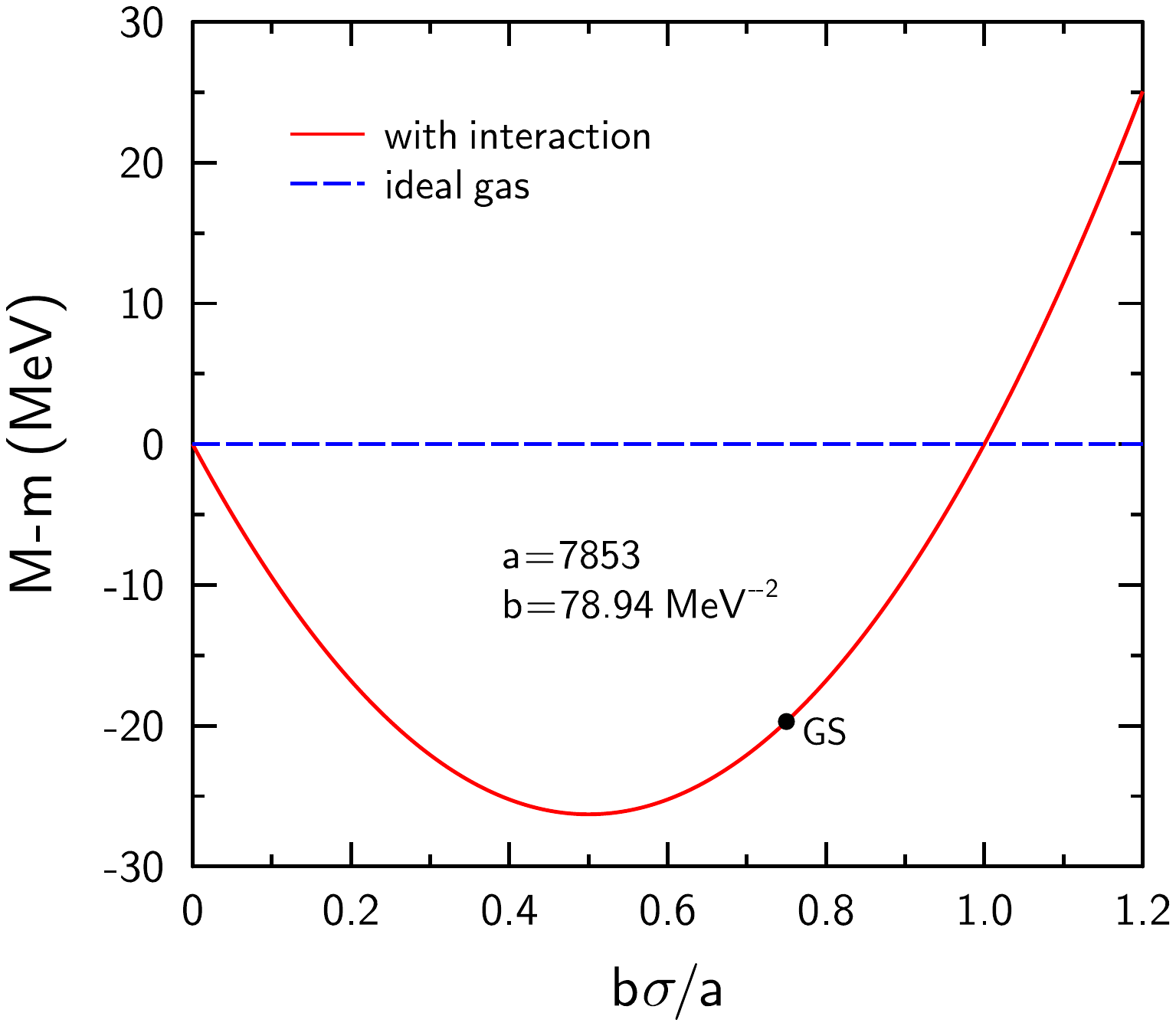}
\caption{
Effective mass of particles in $\alpha$ matter as the function of scalar density $\sigma$. Full dot
shows position of GS\,.
}\label{fig0}
\end{figure}
%------------------------
Substituting (\ref{skin}) into Eqs.~(\ref{efm1}) and (\ref{prex}) gives the explicit relations for the effective
mass and excess pressure\hspm\footnote
{
A similar quasiparticle  model for scalar bosons with
repulsive interaction
was constructed in Ref.~\cite{Sav20}.
}
\begin{eqnarray}
&&M\hspm (\sigma)=\sqrt{m^2-a\hspm\sigma +b\hspm\sigma^2}\,,\label{efm2}\\
&&p_{\,\rm ex}\hspm (\sigma)=
-\frac{a}{4}\hspm\sigma^2 +\frac{b}{3}\hspm\sigma^3\,.\label{epre2}
\end{eqnarray}
Note that $p_{\,\rm ex}$ vanishes at $\sigma=\dfrac{3a}{4b}$ and the minimum value of the effective mass
\mbox{$M_{\textrm{min}}=\sqrt{m^2-\dfrac{a^2}{4b}}$} is~reached at \mbox{$\sigma=\dfrac{a}{2b}$}\,.
A typical plot of $M(\sigma)$ for the case of $\alpha$ matter (\mbox{$m=3727.3~\textrm{MeV}$})
is shown in Fig.~\ref{fig0}. It is seen that at not too high densities the deviation of $M$
from its vacuum value at $\sigma\to 0$ is relatively small.

Similarly to Ref.~\cite{Sat19} we fix the interaction parameters $a,b$ by fitting known
properties of cold $\alpha$ matter. According to microscopic calculations of Ref.~\cite{Cla66}
this matter
has the following values of the binding energy per $\alpha$ particle\hspm\footnote
{
The corresponding binding energy per nucleon, $W=m_N+(W_0-m)/4\approx 12~\textrm{MeV}$ is slightly less than
the binding energy of isosymmetric nuclear matter (about 16 MeV/nucleon).
}
 $W_0$ and the equilibrium density~$n_0$\hspm :
\bel{gsalm}
W_0=m-\left(\frac{\ds\varepsilon}{\ds n}\right)_{\textrm{min}}=19.7~\textrm{MeV}\,,
~~n_0=0.036~\textrm{fm}^{-3}.
\ee

The bosonic matter at $T=0$ consists of condensed
quasiparticles with $\mu=M(\sigma)$, \mbox{$n=n_c=\mu\sigma$}, and
$p=p_{\,\rm ex}(\sigma)$. For the GS one has $d\left(\dfrac{\varepsilon}{n}\right)/dn=p/n^2=0$ and
we obtain the following equations connecting equilibrium parameters $n_0,W_0$ and $\sigma_0$:
\bel{gspar}
p_{\hsp\rm ex}(\sigma_0)=0,~~~m-W_0=M(\sigma_0)=\dfrac{n_0}{\sigma_0}\,.
\ee
As already mentioned, the first equation holds at $\sigma_0=\dfrac{3a}{4b}$. Solving two remaining
equalities, one obtains the following values of parameters $a,b$:
\bel{abval}
a\approx 7853,~~b\approx 78.94~\textrm{MeV}^{-2}\,.
\ee
Note, that the position of GS is indicated  by dot in Fig.~\ref{fig0}.
The GS value $\sigma_0\approx
75~\textrm{MeV}^2$ is significantly lower than the close
packing density $\sigma_p\sim 200~\textrm{MeV}^2$
(it corresponds to the vector density
$n_p\approx m\hspm\sigma_p\sim 0.1~\textrm{fm}^{-3}$~\cite{Sat19}).

It is instructive to calculate the incompressibility modulus:
\bel{incm}
K_0=9\dfrac{dp}{dn}\approx\dfrac{9\hsp p_{\hsp\rm ex}^\prime(\sigma_0)}{m}
=\dfrac{27}{16}\hsp \dfrac{a^2}{b\hsp m}\approx 354~\textrm{MeV}\,.
\ee
This agrees with the
value obtained in Ref.~\cite{Sat19}
for a pure $\alpha$ matter with a mean-field vector potential $U(n)$\,. The latter is
defined as the shift of chemical potential with respect to the ideal gas. One can show
that within the NRA such an approach gives results which are similar to the present model
with scalar interaction. Indeed, at $T=0$ the~chemical potential in the scalar
theory equals $M(\sigma)$ where $\sigma\approx n/m$. Decomposing the~r.h.s.~of \re{efm2}
in powers of $\sigma$ one obtains the Skyrme-like expression\hspm\footnote
{
More exactly, one gets the 'stiff' ($\gamma=1$) version of the potential introduced
in Ref.~\cite{Sat19}\hspm .
}
for the~equivalent vector potential:
\bel{uexp}
U(n)\approx M\hspm (\sigma)-m\approx -A\hspm n+B\hspm n^2,~~
\textrm{where}~~A=\dfrac{a}{2\hspm m^2},~B=\dfrac{b}{2\hspm m^3}\,.
\ee
Substituting the values of $a$ and $b$ from (\ref{abval}),
we get the values of $A,B$ close to those obtained
in Ref.~\cite{Sat19}. However, one can show that the approach with vector potential (\ref{uexp})
leads to superluminal sound velocities $c_s$ at sufficiently large $n$. On the other hand,
the present model does not violate the relativistic
causality condition $c_s<1$ for all   {equilibrium} states.

\section{Phase diagram of $\alpha$ matter}
\label{sec-phase-diagr}

By using the model with vector mean-field interaction
we have calculated in Refs.~\cite{Sat17,Sat19} the phase diagram
of the bosonic matter. It was demonstrated that if this interaction
includes both attractive and repulsive term, the resulting phase diagram
contains both regions of LGPT and BEC. Below we show that a similar phase
diagram takes place in the present model with scalar interaction.

\subsection{The BEC boundary}

As explained above, the BEC boundary in the $(\mu,T)$ plane, $T=T_{\rm BEC}\hspm (\mu)$\,, is
determined by simultaneous solving the equations $M\hspm (\sigma)=\mu, \sigma=\sigma_{\rm th}(T,\mu,\mu)$
where $M$ and $\sigma_{\rm th}$ are given by Eqs.~(\ref{efm2}) and (\ref{gape1}), respectively.
The equation $M\hspm (\sigma)=\mu$ can be solved analytically\hspm\footnote{We have checked that
the second root of this equation corresponds to unstable states.}:
\bel{sbec}
\sigma=\sigma\hspm (\mu)=\dfrac{a+\sqrt{a^2+4\hspm b\hspm (\mu^2-m^2)}}{2\hspm b}\,.
\ee
As shown in Sec.~\ref{sec-bosm}, within the NRA one can use the approximate relations
\mbox{$\sigma_{\rm th}\approx\frac{\ds g~\hspm\xi(3/2)}{\ds \mu\hspm \lambda^3_T(T,\mu)}$}
and $n\approx \mu\hspm\sigma\hspm (\mu)$\,. Finally we get~\re{tbec} for the temperature
$T_{\rm BEC}\hspm (\mu)$\hsp .

\subsection{Liquid-gas phase transition}
Let us consider first the states without BEC. At given $T,\mu$ one can solve Eqs.~(\ref{gape})
and~(\ref{efm2}) with respect to $\sigma$. At sufficiently low temperatures there is a region
of $\mu$ with several solutions $\sigma_i(\mu,T)$. The solution with the largest (lowest) pressure
is thermodynamically stable (unstable)~\cite{Lan75}. The critical line of the LGPT $\mu=\mu\hspm (T)$
is found from the Gibbs condition of phase equilibrium. The latter implies
equality of pressure in coexisting liquid-like ($i=l$) and gas-like ($i=g$) domains
of the mixed phase (MP)\hspm\footnote
{
We neglect surface effects   {associated} with finite sizes of domains.
}.
Within our model one gets three coupled equations for
$\mu\hspm (T),\sigma_g\hspm (T)$, and $\sigma_l\hspm (T)$\,:
\begin{eqnarray}
&&p_{\hsp\rm th}\left[T,\mu,M\hspm (\sigma_g)\right]+p_{\hsp\rm ex}(\sigma_g)=
p_{\hsp\rm th}\left[T,\mu,M\hspm (\sigma_l)\right]+p_{\hsp\rm ex}(\sigma_l)\,,\label{pgibb}\\
&&\sigma_i=\sigma_{\rm th}\left[T,\mu,M\hspm (\sigma_i)\right]~~(i=g,l)\,.~~
\label{sgibb}
\end{eqnarray}
\begin{figure}[htb!]
\centering
\vspace*{-5mm}
\includegraphics[trim=3cm 7.5cm 3cm 7.5cm,width=0.44\textwidth]{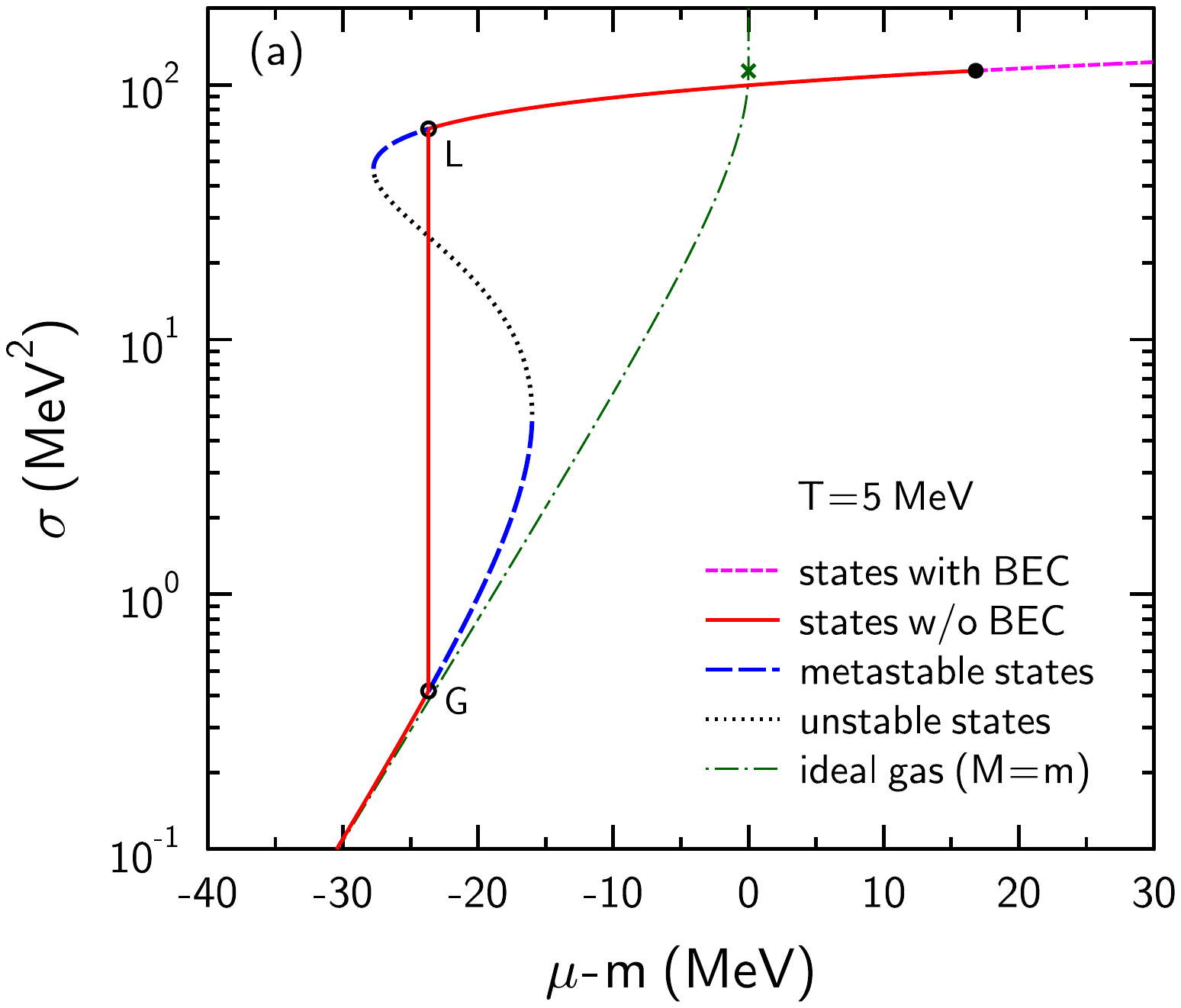}
\includegraphics[trim=2.5cm 7.5cm 3cm 8cm,width=0.455\textwidth]{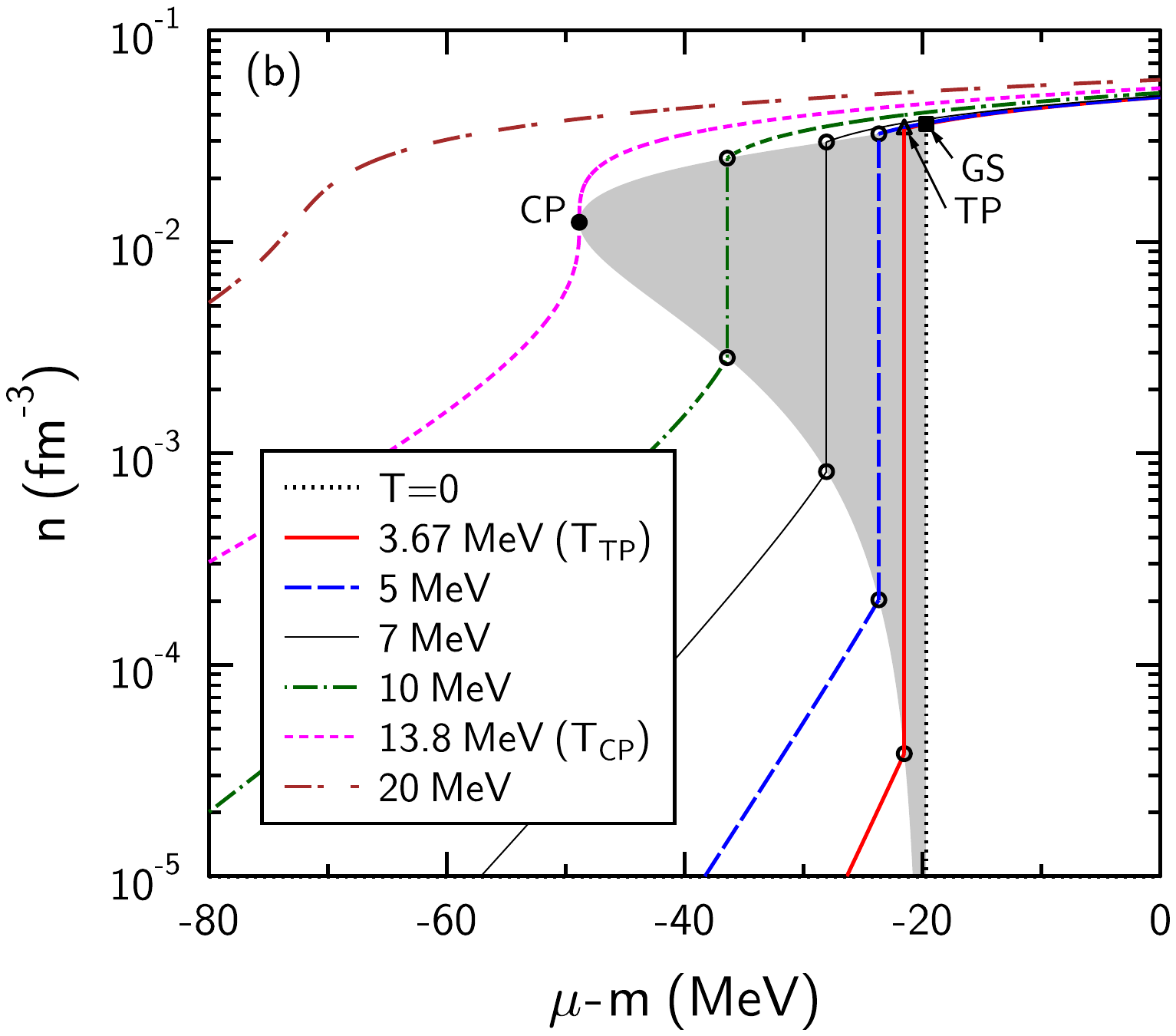}
\caption{
(a) The isotherm $T=5~\textrm{MeV}$ of~$\alpha$ matter on the $(\mu,\sigma)$ plane. The
solid and short-dashed lines show equilibrium states without and with BEC. The filled dot marks
the boundary of the BEC states. The vertical section GL
corresponds to MP states of the LGPT. The dashed and dotted lines represent
the metastable and unstable states. The thin dashed-dotted curve corresponds to the
ideal-gas limit. The cross shows the boundary of BEC states in the
ideal gas. (b) Isotherms of~$\alpha$ matter on the $(\mu,n)$ plane. The filled square, triangle and circle
correspond, respectively, to the GS, TP and CP. The region of MP is shown by shading.
}\label{fig1}
\end{figure}
Solving these equations gives the critical lines $\mu=\mu\hspm (T), p=p\hspm (T)$ of the LGPT
and the values $\sigma_i\hsp (T),M_i\hsp (T),n_i\hsp (T),s_i\hsp (T)\ldots$ at the MP boundaries
('binodals') \mbox{$i=g,l$}\hspm. Note that within the MP region, $\mu$ and $p$ are functions of
temperature only and do not depend, e.g.,~on density. The latter is connected with the volume fraction
of the gas phase~\mbox{$\lambda\equiv V_g/(V_g+V_l)\in [0,1]$}\,:
\bel{vfrac}
n=\lambda\hspm n_g\hspm (T)+(1-\lambda)\hspm n_l\hspm (T)\,.
\ee
At given $T,n$ one gets the relation
\bel{lambda}
\lambda=\frac{n_l\hspm (T)-n}
{n_l\hspm (T)-n_g\hspm (T)}~.
\ee

Our calculations show that at $T=T_{\rm TP}\approx 3.67~\textrm{MeV}$
(so-called
triple point temperature)
the BEC boundary reaches the MP region. At $T<T_{\rm TP}$,
Bose condensate appears in the liquid-phase domains.
In this case
the conditions of phase equilibrium are obtained from~Eqs.~(\ref{pgibb}),~(\ref{sgibb})
after
replacing $M\hspm (\sigma_l)$
by~$\mu$, and $\sigma_l$ by $\sigma\hspm (\mu)$ from~\re{sbec}.

A typical example for $T=5~\textrm{MeV}$ is shown in Fig.~\ref{fig1}~(a).
The solid and short-dashed lines show equilibrium states on the $(\mu,\sigma)$ plane.
At the considered temperature the BEC states lie outside the MP region
(the vertical line GL). One can see that at large scalar densities the results strongly deviate
from the ideal gas of bosons with the vacuum mass $m$\,.
We also show positions of metastable and unstable (spinodal) states. They correspond
to solutions of the gap equation with lower values of pressure
as compared to equilibrium state at the same $T$ and $\mu$\,.

\begin{table}[ht!]
\caption
{\label{tab1} Characteristics of critical point of $\alpha$ matter}
\vspace*{3mm}
\begin{tabular}{|c|c|c|c|c|c|c|c|}
\hline
~$T~(\textrm{MeV})$~&~$\mu-m~(\textrm{MeV})$~&~$M-m~(\textrm{MeV})$~&~$\sigma~(\textrm{MeV}^2)$~&
~\mbox{$n~\left(\textrm{fm}^{-3}\right)$}~&~~$S/N$~~&~~$p/(nT)$~~&~~$n\lambda^3_T$~~\\
\hline
~$13.8$&$-48.9$&$-20.0$&$25.8$&$0.0124$&$4.54$&$0.334$&$0.129$\\
\hline
\end{tabular}
\end{table}
In Fig.~\ref{fig1}~(b) we compare different isotherms of $\alpha$ matter on the $(\mu,n)$ plane.
The lower and upper boundaries of the MP (shaded region) correspond, respectively to the
gas-like and liquid-like binodals. Note that at $T=0$ the liquid-like binodal state
coincides
with the~GS of equilibrium $\alpha$ matter.
One can see that the LGPT disappears when $T$ exceeds the critical point (CP)
temperature $T_{\rm CP}\approx 13.8~\textrm{MeV}$. At this point
the derivative $(\partial\hspm n/\partial\mu)_T$ diverges. Table I shows characteristics
of CP obtained within the~present model.

\begin{figure}[htb!]
\centering
\includegraphics[trim=2cm 8cm 3cm 8cm,width=0.6\textwidth]{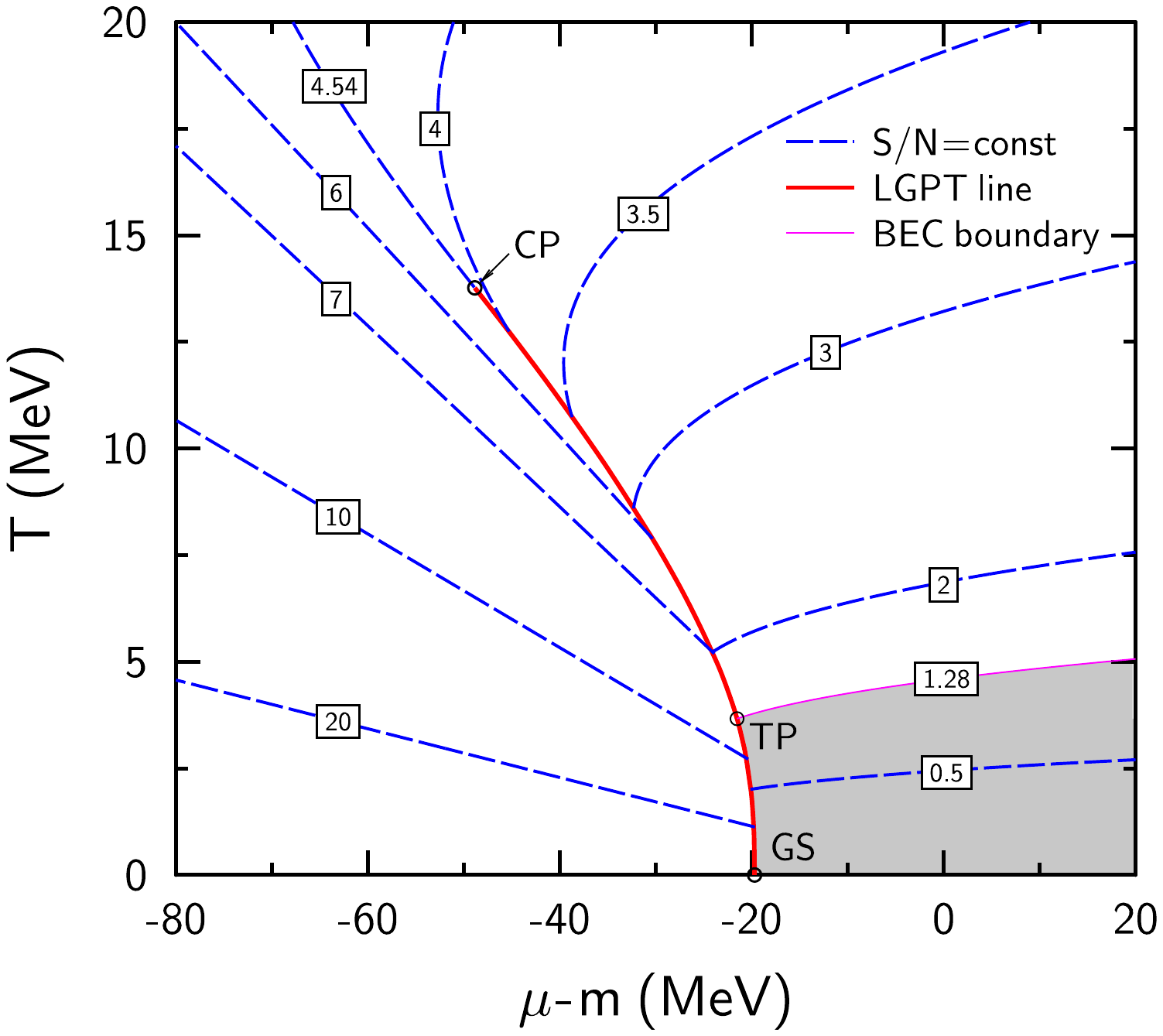}
\caption{
Phase diagram of $\alpha$ matter on the $(\mu,T)$ plane. The dashed curves are adiabatic trajec\-tories,
i.e.,~lines of constant entropy per particle. The values of $S/N$ are given in boxes. The~{BEC region} is shown by shading.
}\label{fig3}
\end{figure}
Figure~\ref{fig3} represents the phase diagram on
the
$(\mu,T)$ plane. The MP region corresponds to
the thick solid line between the GS and the CP.
The BEC states lie below the thin curve which crosses
the LGPT critical line at the triple point~TP.
The domain of BEC states is shown in Fig.~\ref{fig3}
by the shaded area.
At the same plot we show the behavior of adiabatic
trajectories (isentropes), i.e., the lines of constant specific entropy \mbox{$S/N=\textrm{const}$}.

The isentropes play an important role in a fluid-dynamical evolution of excited matter\hspm\footnote{In absence
of dissipation and shock waves, the total entropy is conserved for thermally equilibrated matter even in presence of collective flow.}.
In particular, we would like to mention the hydrodynamic~\cite{Lan53} and thermal~\mbox{\cite{Cle98,And06}}
\mbox{models} of heavy-ion collisions which successfully describe particle production at high energies. They postulate that a hot
and compressed 'fireball' is formed at some intermediate stage of
a~nuclear collision. Due to the presence of internal
pressure, the fireball expands at later stages,
producing secondary particles.
Fitting the observed data confirmed that the specific entropy of fireball is fully determined by the initial~c.m. energy of colliding nuclei.

As one can see in Fig.~\ref{fig3}, the isentropes enter the MP region at any $S/N$.
Especially important is the isentrope $S/N=(S/N)_{\,\rm CP}\approx 4.54$
which goes through the~CP. As will be shown later, the trajectories with the specific entropy close to $(S/N)_{\,\rm CP}$ go through
states with anomalously large fluctuations of the particle density. Note, that isentropes with~$S/N$ larger (smaller) than $(S/N)_{\,\rm CP}$
enter the MP region at the gas-like (liquid-like) side.
As mentioned above,
the isentropes do not cross the BEC boundary outside the~MP region. We shall come back to discussing
this phase diagram in Sec.~\ref{sec-fluct} (see Fig.~\ref{fig9}).

\begin{figure}[htb!]
\centering
\includegraphics[width=0.7\textwidth]{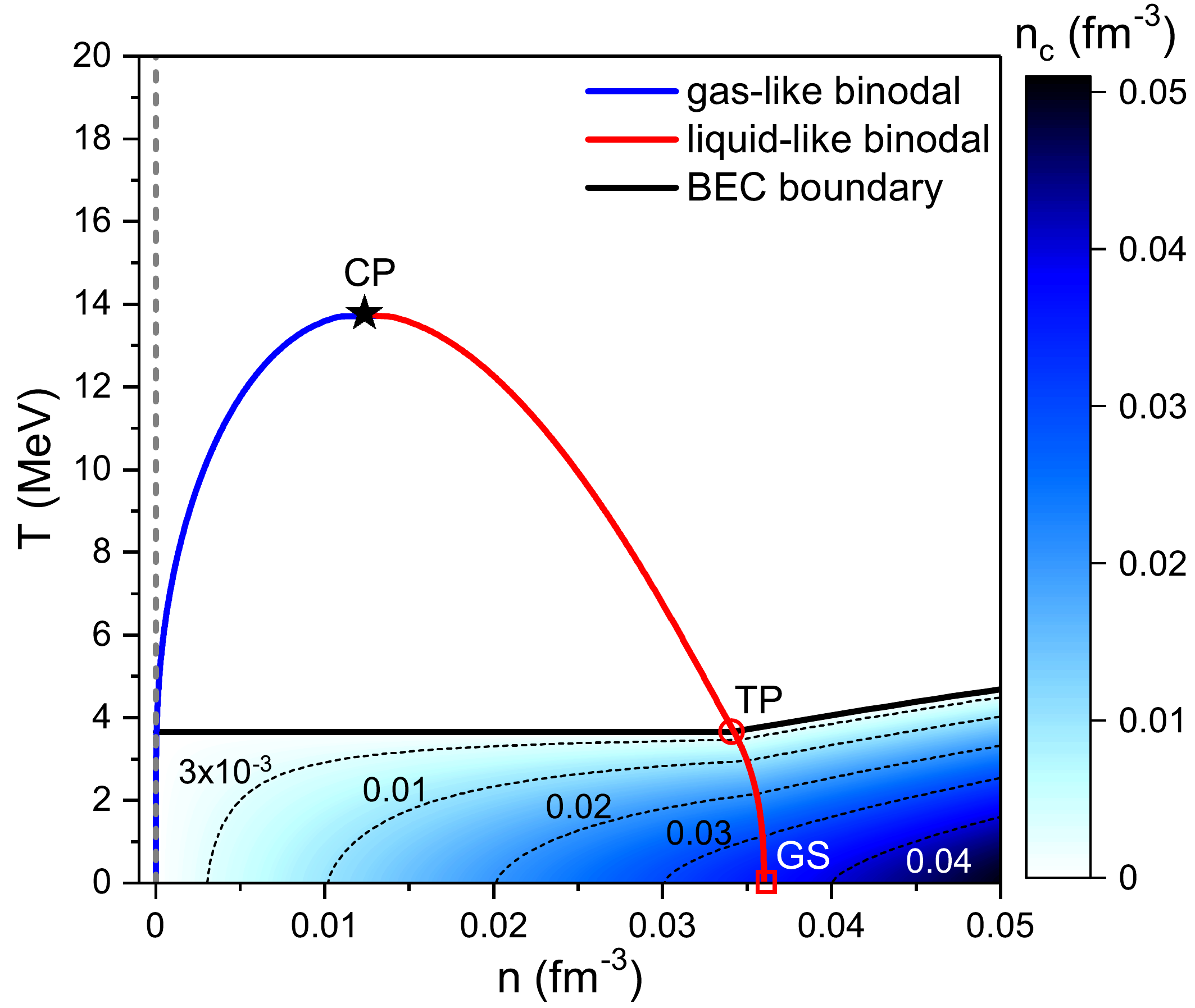}
\caption{
Phase diagram of $\alpha$ matter on the $(n,T)$ plane. The values of the condensate density~$n_c$
are shown by different shades of blue color (see the color map on r.h.s.). The dotted lines
are contours of equal $n_c$ in the BEC region. The square, circle and star correspond, respectively,
to the GS, TP, and CP.
}\label{fig4}
\end{figure}
Let us now calculate explicitly the condensate density $n_c$ at fixed $n,T$,
first, for states outside the MP, i.e., for $n>n_l\hspm (T)$ and $T<T_{\rm BEC}$.
As explained in Sec.~\ref{sec-bosm} this can be done by solving the equations
\bel{cdvn}
n_c=\mu\left[\sigma\hspm (\mu)-\sigma_{\hspm\rm th}(T,\mu,\mu)\right]=n-n_{\hspm\rm th}(T,\mu,\mu)\,,
\ee
where $\sigma\hspm (\mu)$ is given by~\re{sbec}. One can see that $n_c$ increases from zero
to $n$ when temperature drops from $T=T_{\rm BEC}$ to zero.

\begin{figure}[htb!]
\centering
\includegraphics[trim=2cm 8cm 3cm 8cm,width=0.48\textwidth]{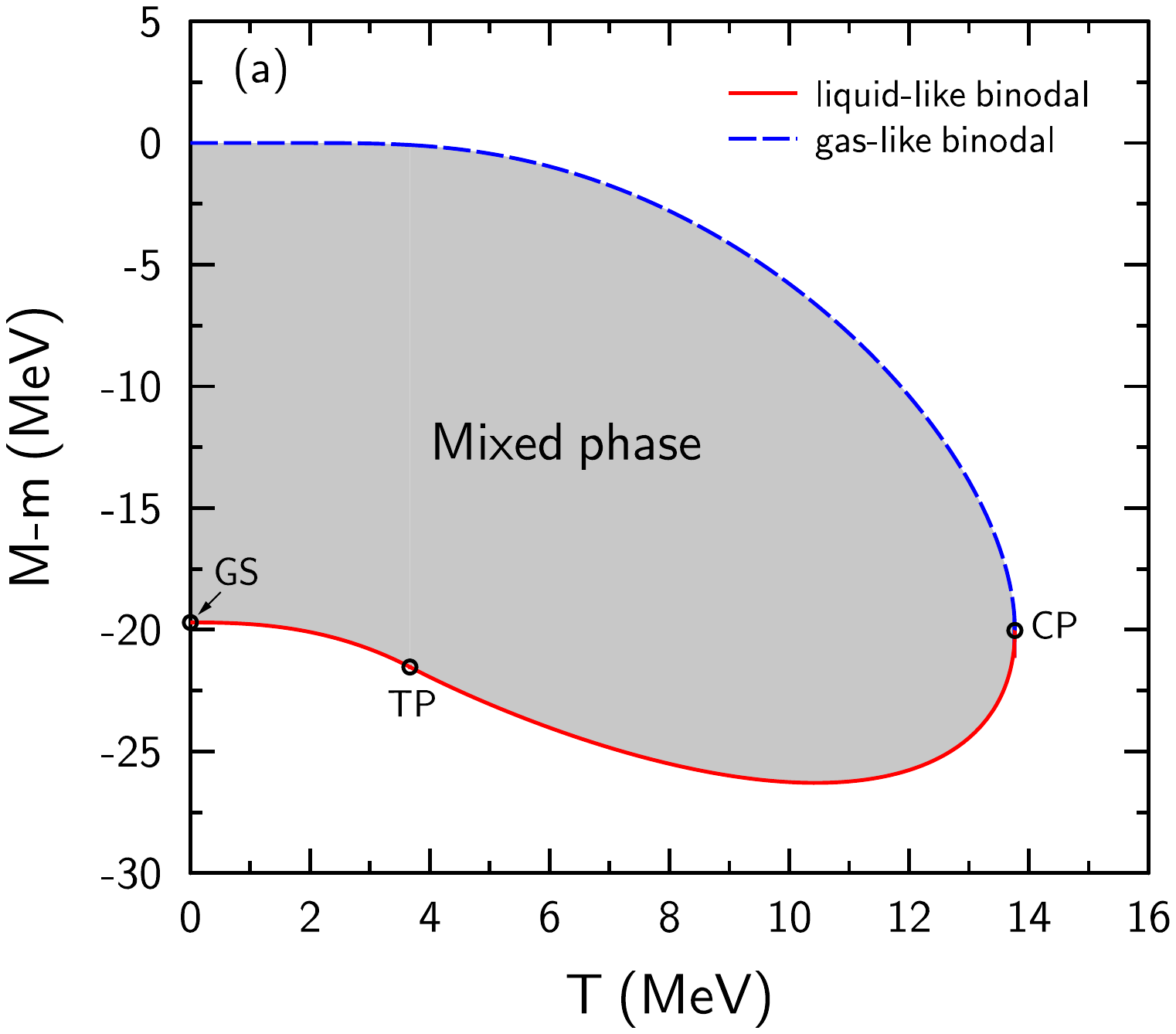}
\includegraphics[trim=2cm 8cm 3cm 8cm,width=0.48\textwidth]{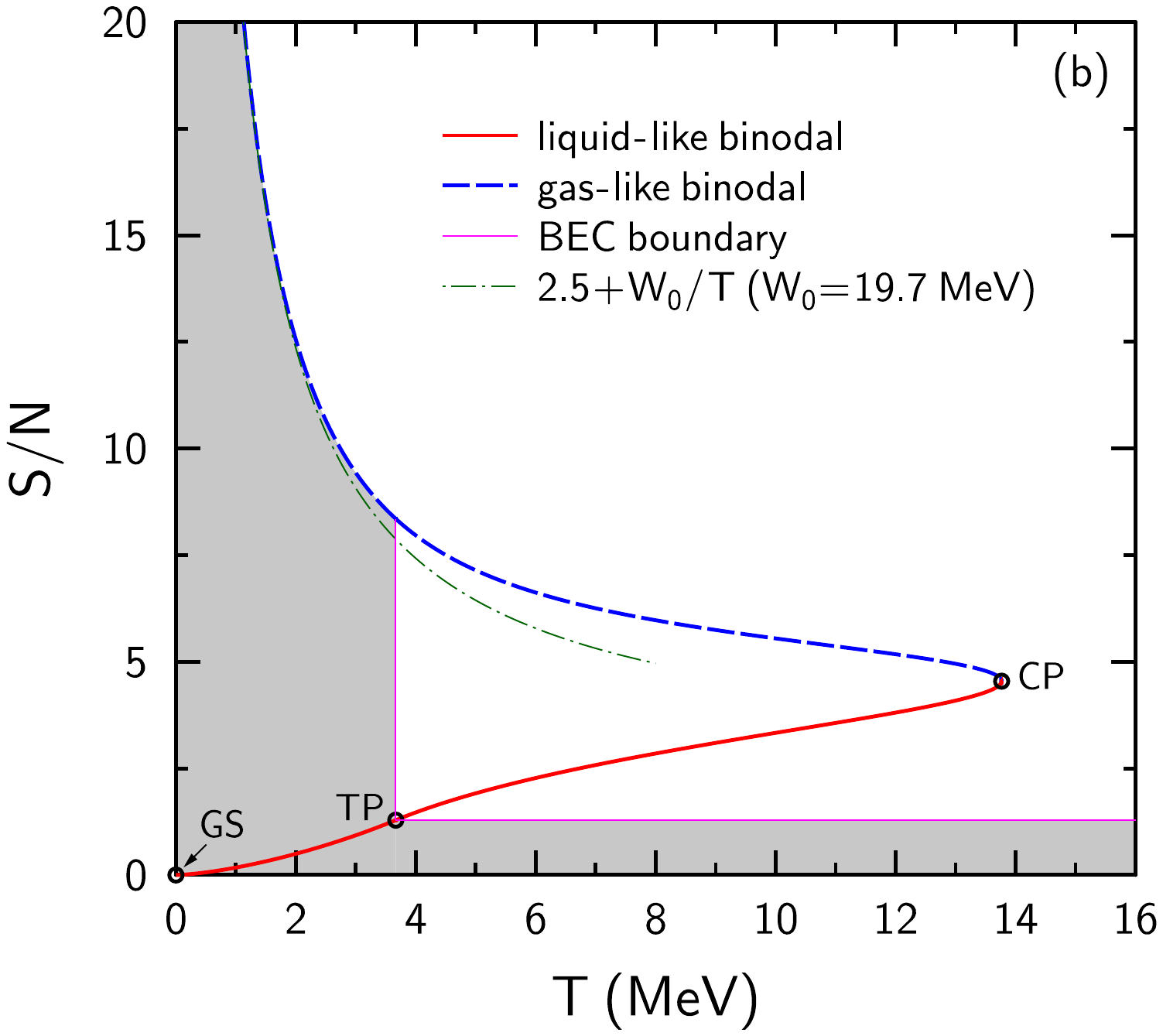}
\caption{
 {Phase diagrams} of $\alpha$ matter on the ($T,M$) (a) and ($T,S/N$) (b) planes.
The regions of~MP (a) and BEC (b) are shown by shading.
}\label{fig5}
\end{figure}
Inside the MP the condensate appears at $T<T_{\rm TP}$. At such temperatures
the densities of the gas- and liquid-like domains satisfy the relation $n_g(T)\ll n_l(T)$\,.
Unless the temperature is extremely low, one can disregard presence of BEC in the gaseous phase.
Then one can write the relation \mbox{$\langle n_{c}\rangle\approx (1-\lambda)\, n_{cl}(T)$}
for the density of BEC averaged over the ensemble of coexisting domains.
Here $n_{cl}(T)$ is the condensate density in the liquid phase at \mbox{$T<T_{\rm TP}$}.
Substituting further~\re{lambda} and
neglecting terms $\sim n_g/n_l$ one gets the~relation
\mbox{$\langle n_{c}\rangle/n\approx n_{cl}(T)/n_l(T)$}.
More detailed information is given in Fig.~\ref{fig4} where we show lines of
equal condensate densities on the $(n,T)$ plane. The obtained phase diagrams are similar
to those derived in~Ref.~\cite{Sat17}. It is interesting that they qualitatively agree
with phase diagrams observed~\cite{Buc90} for atomic $^4$He.
\begin{figure}[htb!]
\centering
\includegraphics[trim=2cm 7.5cm 3cm 8.5cm,width=0.6\textwidth]{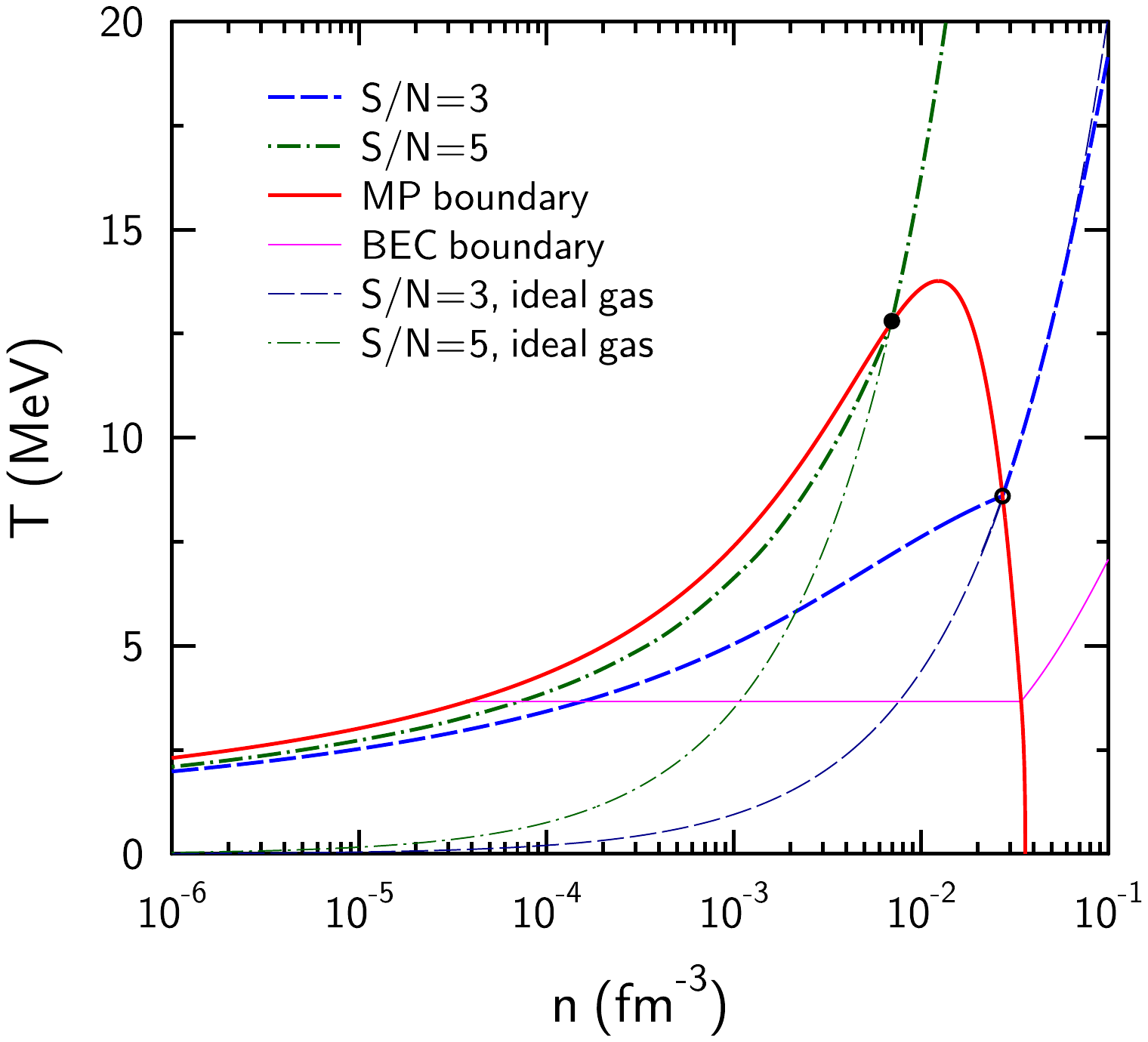}
\caption{
The   {isentropes} $S/N=3$ (the dashed line) and $S/N=5$ (the dash-dotted line) on the ($n,T$) plane.
The dots mark their intersection with the MP boundary. Thin dashed and dash-dotted lines represent the ideal-gas calculation.
}\label{fig6}
\end{figure}

Figure~\ref{fig5}\hsp (a) shows the phase diagram on the $(T,M)$ plane. One can see
that~\mbox{$M_g(T)>M_l(T)$}\,. At low temperatures $M_g(T)\approx m$
and $M_l(T)=\mu\hsp (T)\approx m-W_0$
[see~\re{gspar}]. In Fig.~\ref{fig5}\hsp (b) we show the values of specific entropy $\widetilde{s}_i=(S/N)_i$ at
the MP boundaries $i=g,l$. One can see that
$\widetilde{s}_l(T)<\widetilde{s}_{\rm CP}<\widetilde{s}_g(T)$\,. It is possible to derive
a simple analytic formula for $\widetilde{s}_g(T)$ at $T\to 0$. Indeed, in this limit the density $n=n_g(T)$ is small and
one can apply the Boltzmann approximation in calculating the specific entropy on the gas-like binodal. One gets approximate relations
\bel{sentg}
\widetilde{s}_g(T)\approx\frac{5}{2}-\ln{z}\approx\frac{5}{2}+\frac{m-\mu\hspm (T)}{T}
\approx\frac{5}{2}+\frac{W_0}{T}\,.
\ee
In the first equality we have used the upper line of~\re{spent}.
Comparison of the dashed and dash-dotted lines in Fig.~\ref{fig5}\hsp (b) confirms
good accuracy of these relations at low $T$\,.

In Fig.~\ref{fig6} we compare the behavior of isentropes $S/N=3$ and $S/N=5$ on the ($n,T$) plane.
Within the MP region we use the relation for the entropy density
\bel{entd2}
s=\lambda\hsp s_g(T)+(1-\lambda)\hsp s_l(T)\hsp,
\ee
where $\lambda$ is given by~\re{lambda}\,.
One can see jumps in slopes of isentropes at the MP
boundary (a similar behavior takes place also on the $(n,p)$ plane). This in turn leads to
jumps of the sound velocity and the heat capacity at these boundaries (see Sec.~\ref{sec-cs} and~\ref{sec-cv}).
Note that at small $n$ and $T$ isentropes go near the gas-like binodal.
The calculation shows that outside the MP region the adiabatic trajectories are close to
those in the ideal Bose-gas.

\section{Possible signatures of phase transition in {\large $\alpha$} matter}

\subsection{Strong density fluctuations}
\label{sec-fluct}

Statistical fluctuations of conserved charges (e.g., baryon number)
are important observables for
experimental studies of the phase diagrams of interacting systems.
In the grand-canonical ensemble
such fluctuations are expressed via 'susceptibilities', i.e., higher-order derivatives of
pressure with respect to $T,\mu$.
Below we calculate the scaled
variance $\omega$ defined as the second moment of the particle number in a given volume:
\bel{omdef}
\omega=\dfrac{\langle N^2 \rangle-\langle N \rangle^2}{\langle N \rangle}\,,
\ee
where averaging is performed at fixed
temperature.
Outside the MP region one gets~\cite{Lan75} the relation
\bel{omder}
\omega=\frac{T}{n}\left(\frac{\partial\hspm n}{\partial \mu}\right)_T=
\frac{T}{n}\left(\frac{\partial^{\,2}\hspm p}{\partial \mu^{2}}\right)_T.
\ee
In calculating the derivatives in the r.h.s.
of~(\ref{omder}) one should explicitly take into account the dependence $M=M\hspm (T,\mu)$\,.

At the region $T>T_{\rm BEC}$, using Eqs.~(\ref{vden}), (\ref{efm2}) and (\ref{gape}), one has
\bel{ome1}
\omega=
\frac{T}{n}\left[\frac{\partial\hspm  n_{\rm th}(T,\mu,M)}{\partial \mu}+\frac{\partial\hspm
n_{\rm th}(T,\mu,M)}{\partial M}\hsp  M^\prime (\sigma)\hsp\left(\frac{\partial \sigma}{\partial \mu}\right)_T\right],
\ee
where
\bel{sig4}
M'(\sigma)=\left(b\hspm \sigma -\dfrac{a}{2}\right)\hspm M^{-1},~~~\left(\frac{\partial \sigma}{\partial \mu}\right)_T=
\frac{\partial \sigma_{\rm th}}{\partial \mu} \left[1-\frac{\partial \sigma_{\rm th}}{\partial M}\hsp M^\prime (\sigma)
\right]^{-1}.
\ee
At \mbox{$\mu\to\mu_{\hspm\rm CP}, T\to T_{\rm CP}$} both $(\partial\hspm n/\partial\mu)_T$ and $(\partial\hspm\sigma/\partial\mu)_T$
diverge which leads to the~relation
\bel{crpc}
\frac{\partial \sigma_{\rm th}}{\partial M}\,M^\prime (\sigma)=1~~~\textrm{at CP}.
\ee
As one can see from~\re{gape1} the density $\sigma_{\rm th}$ decreases with $M$ at fixed $T,\mu$. Therefore, the~CP
(and LGPT) may exist if $M\hspm (\sigma)$ contain regions with $M^\prime\hspm (\sigma)<0$. In accordance with first equality in~\re{sig4}, this is possible
only if the interaction has an attractive term with nonzero $a$.

By using Eqs.~(\ref{snnr}) and (\ref{sntr1}) one can calculate the density derivatives
entering \re{ome1} in the NRA. Then
one obtains
\bel{sig4a}
\left(\frac{\partial \sigma}{\partial\mu}\right)_T\approx \dfrac{\sigma\hspm g_{1/2}(z)}{\sigma\hspm M^\prime (\sigma)g_{1/2}(z)+Tg_{3/2}(z)}\,.
\ee
At given $n,T$, the values $z$ and $\sigma$ are determined from
the approximate rela\-tions 
\bel{sig4b}
n\approx m\sigma\approx g\hsp g_{3/2}(z)/\lambda_T^3(T,m)\,.
\ee
Substituting (\ref{sig4a}) into
(\ref{ome1}) gives\hspm\footnote
{
A similar result can be obtained in the vector model of Ref.~\cite{Sat17} with the replacement
$\sigma M'(\sigma)\to n\hsp U'(n)$ where $U(n)$ is the potential introduced in~\re{uexp}\,.
}
\bel{ome1a}
\omega\approx\left[\dfrac{g_{3/2}(z)}{g_{1/2}(z)}+\dfrac{\sigma M^\prime (\sigma)}{T}\right]^{-1}.
\ee
The second term in denominator vanishes
in the limiting case of the ideal Bose gas ($M=m$).
Note, that at $T\to T_{\rm BEC}$ the first term in brackets goes to zero.

At $T<T_{\rm BEC}$ using Eq.~(\ref{condt}) one has:
\bel{ome2}
\omega=
\frac{T}{n}\left[\frac{\partial\hspm n_{\hspm\rm th}(T,\mu,\mu)}{\partial \mu}+\frac{\partial\hspm n_c(T,\mu)}{\partial \mu}\right],
\ee
where, $n_c(T,\mu)$ is defined by the first equality in (\ref{cdvn}). Within the NRA one has approximately
\mbox{$n\approx\mu\hspm\sigma\hspm (\mu)$}, where $\sigma\hspm (\mu)$ is given by~\re{sbec}\,. Using further
\re{omder} one gets
the relations
\bel{ombec}
\omega\approx\frac{T\sigma^\prime(\mu)}{\sigma(\mu)}=\frac{\mu T}{\sigma\hsp (b\hspm\sigma-0.5\hspm a)}\,.
\ee
One can see that $\omega\to 0$
at $T\to 0$\,. This greatly deviates from the ideal-gas case~\mbox{($a=0, b=0$)} where~$\omega$ is infinite~\cite{Lan75} in the BEC region.

%----------------------
\begin{figure}[htb!]
\includegraphics[trim=2cm 8cm 3cm 8.5cm,width=0.48\textwidth]{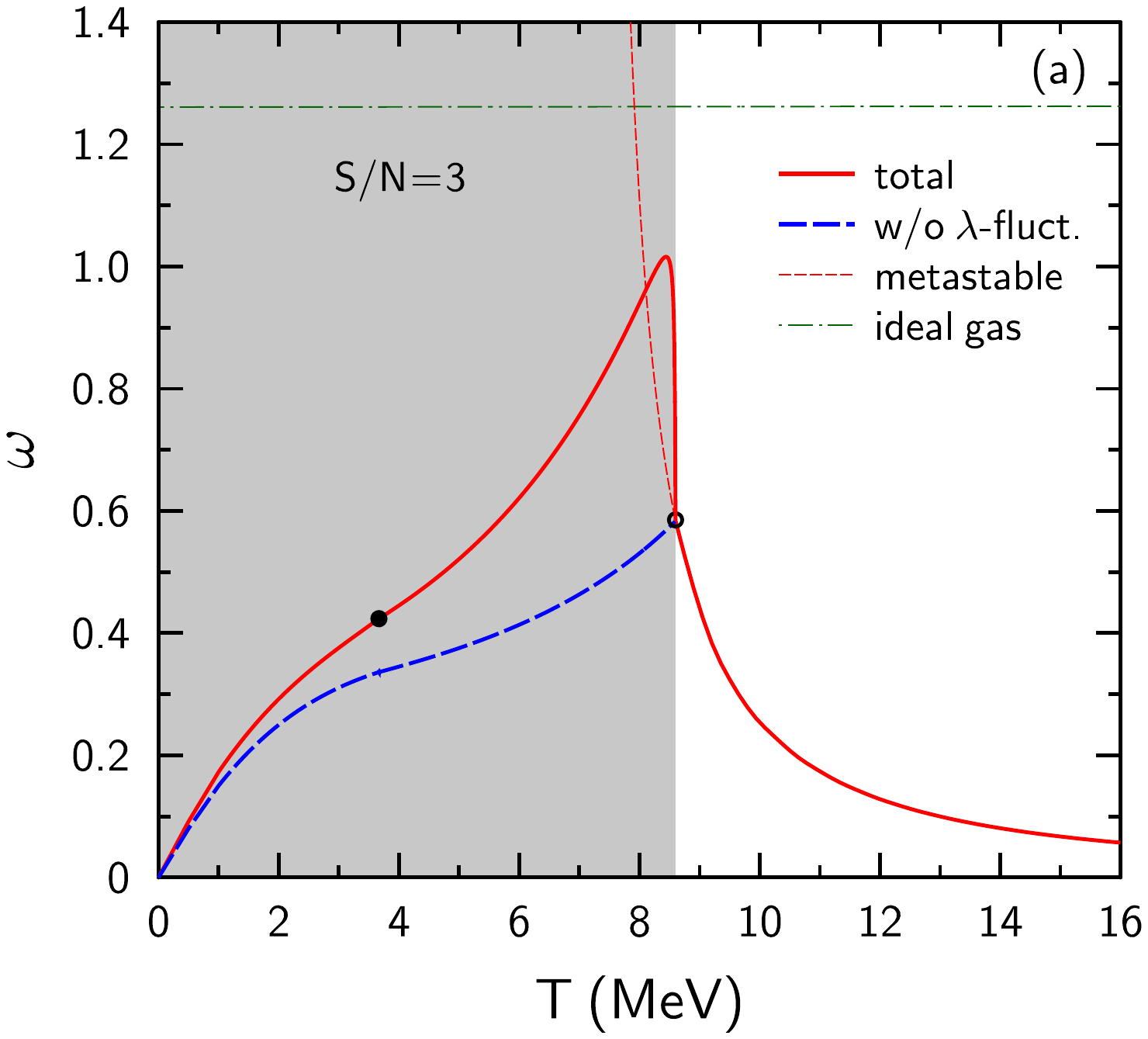}
\includegraphics[trim=2cm 8cm 3cm 8.5cm,width=0.48\textwidth]{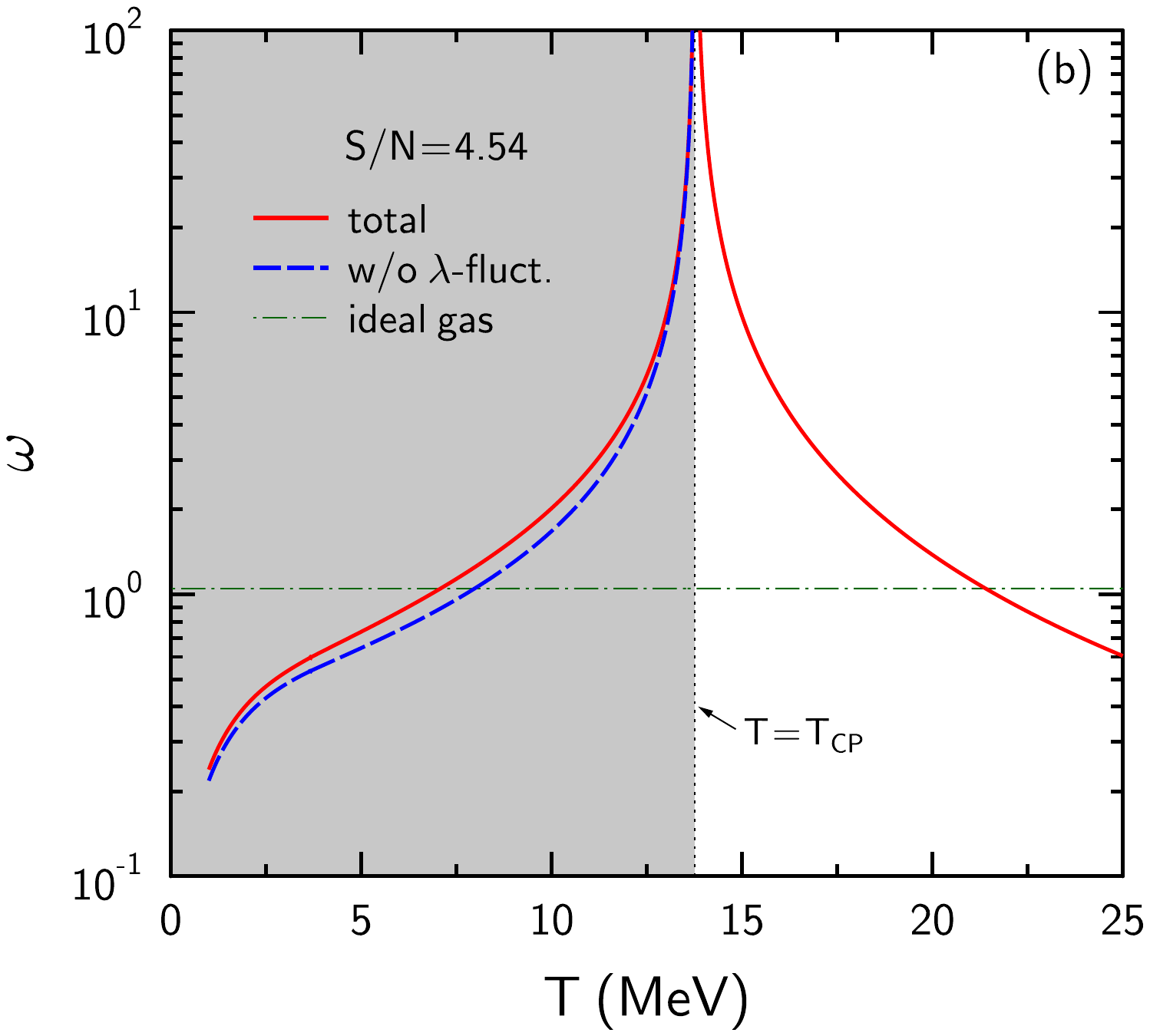}
\caption{
Scaled variance of $\alpha$ matter as the function of temperature along the isentropes \mbox{$S/N=3$~(a)}
and  \mbox{$S/N=4.54$~(b)}. The dashed lines are calculated without the third term in \re{omega-mp}. The MP
regions are shown by shading. The open and full dots correspond, respectively, to boundaries of MP and BEC states.
}\label{fig7}
\end{figure}
\begin{figure}[htb!]
\centering
\includegraphics[trim=2cm 8cm 3cm 8.5cm,width=0.6\textwidth]{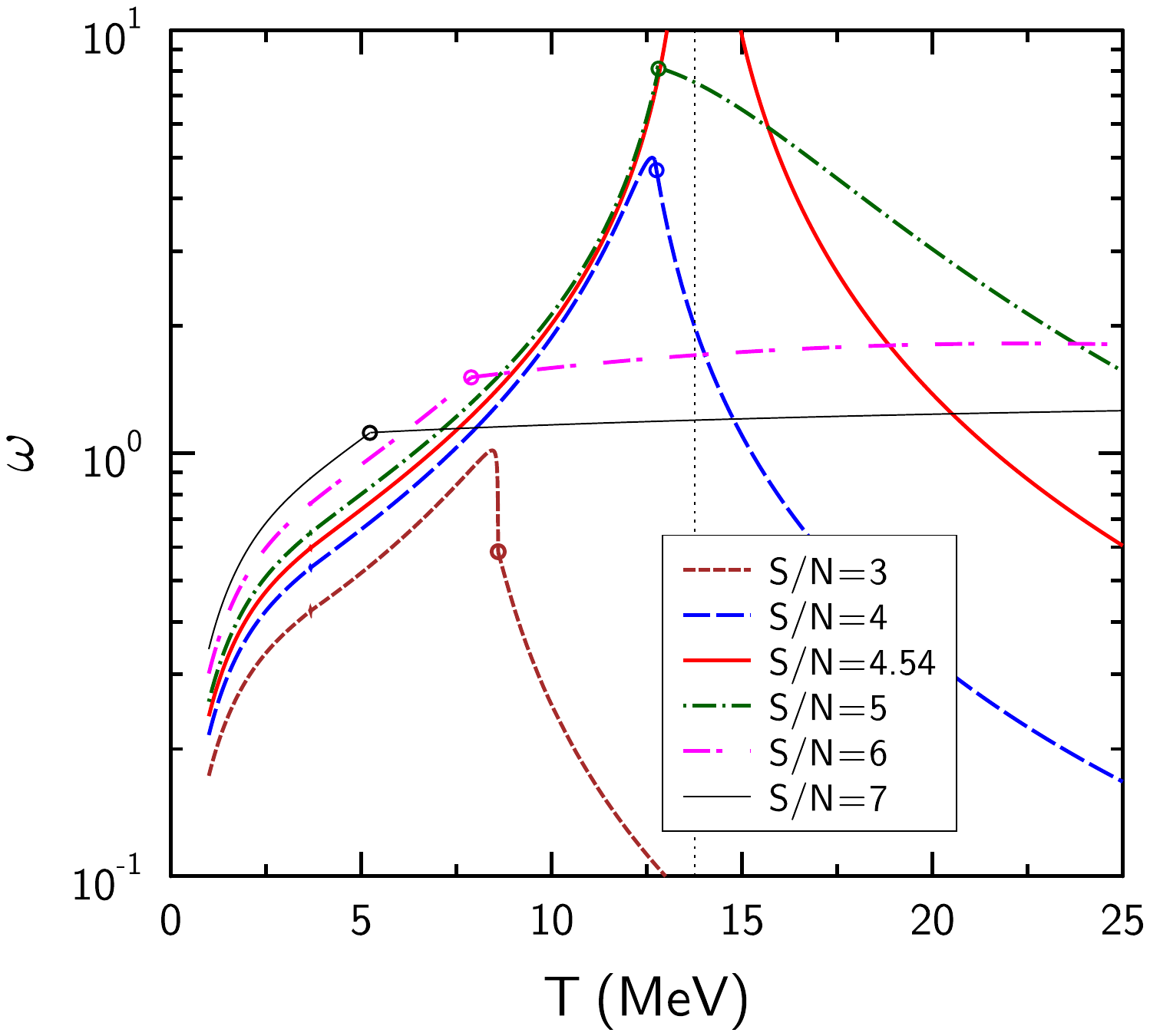}
\caption{
Scaled variance of $\alpha$ matter as the function of temperature along the isentropes
with \mbox{$S/N=3-7$}. Circles mark points where the isentropes cross the MP boundary.
  {The vertical dotted line shows the temperature $T=T_{\rm CP}$\hspm .}
}\label{fig8}
\end{figure}
%------------------------------
\begin{figure}[htb!]
\centering
\includegraphics[width=\textwidth]{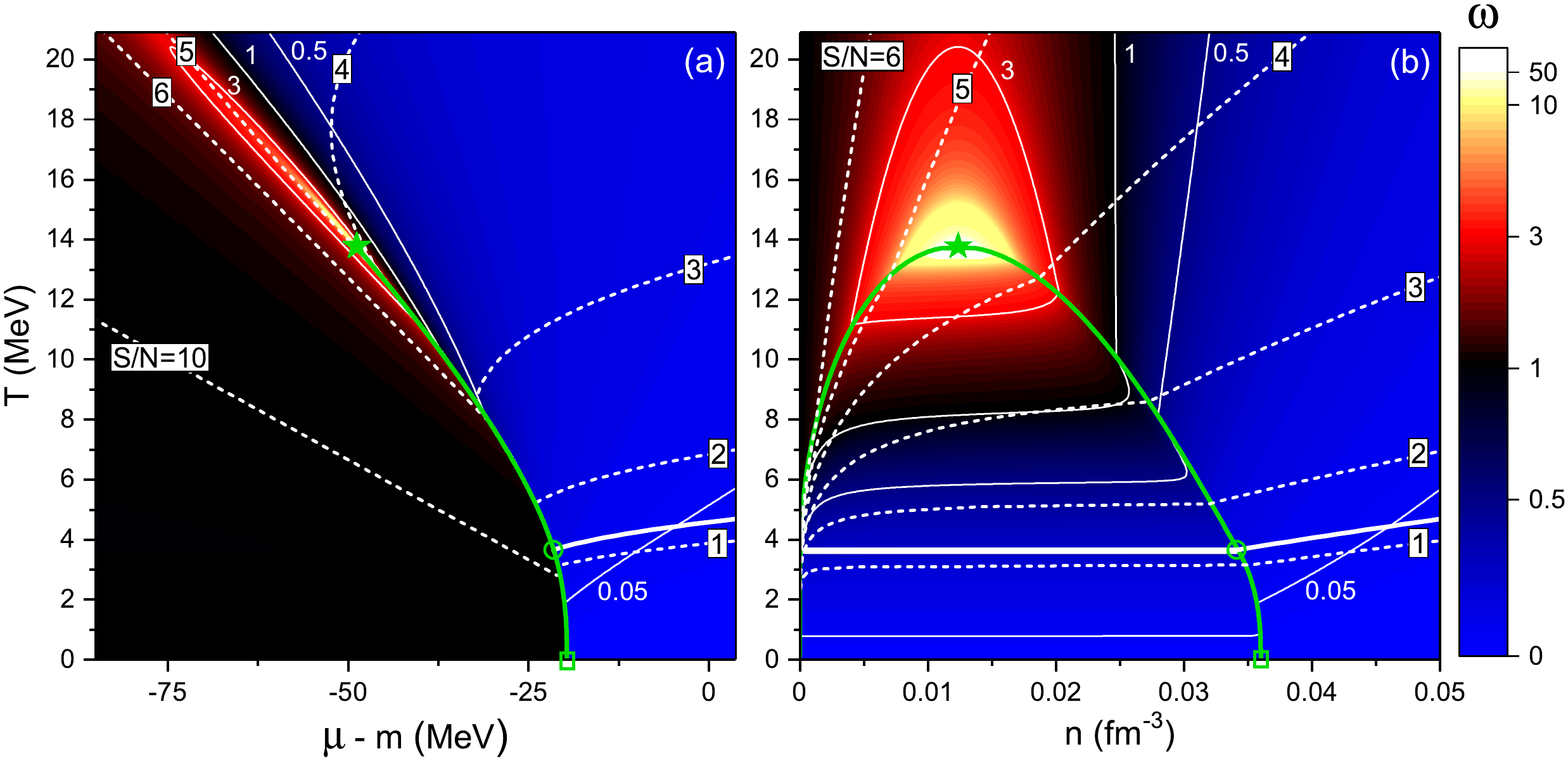}
\caption{
{(a) The dashed lines: isentropic trajectories with different $S/N$ (shown by black numbers in boxes)}
on the $(\mu,T)$ plane. Colors show values of scaled variance $\omega$\hspm .   {Thin white lines are contours of equal
$\omega$\, (their values are given by white numbers). The thick green and white lines represent the LGPT line and the
BEC boundary, respectively. The star marks position of the CP.}
(b) Same as (a) but on the $(n,T)$ plane. Note the strong $\omega$ peak near the critical point.
}\label{fig9}
\end{figure}
%-----------------------
In the MP region one should take into account not only fluctuations of particle numbers inside the coexisting domains of matter,
but also fluctuations of the interphase boundaries, which change the relative fraction of domain volumes. The latter corresponds
to fluctuations of the parameter $\lambda$ around the equilibrium value given by~\re{lambda}. One can include both types of fluctuations
by calculating $\omega$ directly from \re{omdef}\,.
In Ref.~\cite{Vov15} such a calculation was made for a particular case of the classical van der Waals model. Below we apply
a~similar approach for an arbitrary LGPT\hspm\footnote{
Higher-order particle number fluctuations can be calculated by using the method
developed recently in~Ref.~\cite{Pob20}.
}.

We obtain the following result at given $n,T$ in the MP region:
\bel{omega-mp}
\omega=\frac{1}{n}\left[\lambda\hspm n_g\omega_g
+(1-\lambda)\hspm n_l\hspm\omega_l+(n_l-n_g)^2\hsp\frac{\lambda\hspm (1-\lambda)\hspm\omega_g\hspm \omega_l}
{\lambda\hspm n_l\hspm\omega_g+ (1-\lambda)\hspm n_g \hspm\omega_l}\right].
\ee
Here $\lambda$ is defined in~\re{lambda}, $n_i$ and $\omega_i$ are, respectively, the particle densities and scaled variances
inside the domains $i=g,l$. These quantities are equal to their values at the gas- and liquid-like binodals.
The contribution of $\lambda$ fluctuations
is described by the third term in the r.h.s. of~(\ref{omega-mp}). This term
vanishes near the MP boundary where $\lambda\hspm (1-\lambda)\to 0$.

The results of $\omega$ calculation for states along the isentropes with different
$S/N$ are shown in Figs.~\ref{fig7}--\ref{fig8}. The calculations predict strong peaks of scaled variance for $S/N=4.5\pm 1$\,.
In Fig.~\ref{fig7} we make the comparison with the ideal-gas calculation. One can see the strong sensitivity of~$\omega$ to particle
interactions. Figures~\ref{fig9}\hsp (a) and (b) show the density plots of $\omega$ in the $(\mu,T)$ and $(n,T)$ planes, respectively.
One can see a very narrow peak of $\omega(T,\mu)$ at $T\approx T_{\rm CP}$\,. This can be used as a clear signal of the CP
in experimental searches of LGPT.

\subsection{Softening of the equation of state}
\label{sec-cs}

An important characteristics of the equation of state is the
sound velocity $c_s$ which characterizes propagation of a small perturbation in the local rest frame of matter\,. In the ideal
fluid dynamics the sound velocity squared is equal to~\cite{Lan75}
\bel{souv}
c_s^2=\left(\dfrac{\partial\hspm p}{\partial\varepsilon}\right)_{\widetilde{s}}=
\left(\dfrac{\partial\hspm p}{\partial\varepsilon}\right)_n+\dfrac{n}{\varepsilon+p}
\left(\dfrac{\partial\hspm p}{\partial n}\right)_\varepsilon\,,
\ee
where $\widetilde{s}=s/n$ is the entropy per particle\hspm\footnote
{
Equation~(\ref{souv}) is derived for a continuous matter without
large gradients of density. On the other hand, this is not true for MP with
different densities of liquid and gas
domains. Nevertheless, one can approximately consider the MP region
as a homogeneous matter if the wavelength of a sound wave
exceeds typical domain sizes (i.e., at low enough
frequencies).
}.
Using~Eqs.~(\ref{dpre}),~(\ref{deps}) one can directly calculate~$c_s$ by expressing the derivatives entering ~\re{souv} via
the derivatives of $n,s$ over~$T,\mu$\,.
For the MP states one can use Eqs.~(\ref{vfrac}),~(\ref{entd2}) and rewrite these derivatives via the binodal
quanti\-ties~$s_i^{\hsp\prime}\hspm(T)$ and $n_i^\prime\hspm (T)~(i=g,l)$\,.

Within the NRA one can get much simpler expressions (outside the MP).   Indeed, from the first equality in (\ref{souv}), substituting
$\varepsilon\approx m\hspm n$
one has
\bel{cs2a}
c_s^2\approx \frac{1}{m}\left[\left(\dfrac{\partial\hspm p_{\hspm\rm th}}{\partial n}\right)_{\widetilde{s}}+\left(\dfrac{\partial\hspm p_{\rm ex}}
{\partial n}\right)_{\widetilde{s}}\right].
\ee
According to \re{spent}, in this approximation the specific entropy
$\widetilde{s}$ depends in the NRA only on fugacity $z$ at $T>T_{\rm BEC}$ or on $n\lambda_T^3$
at $T<T_{\rm BEC}$. From Eqs. (\ref{snnr}) and (\ref{penr}) one can see that in both cases $n\hspm T^{-2/3}$
and $p_{\hspm\rm th}\hspm T^{-5/3}$
are approximately constant at $\widetilde{s}=\textrm{const}$.
Using~\re{prex1} one gets
\bel{cs2b}
c_s^2\approx\dfrac{\sigma\hspm M^{\prime}(\sigma)}{m}+
\frac{5T}{3m}\,
\left\{\begin{array}{ll}
\dfrac{g_{\,5/2}\hspm (z)}
{g_{\,3/2}\hspm (z)},~~~~~~&T>T_{\rm BEC}\,,\\[3mm]
\dfrac{\xi\hspm (5/2)}
{n\lambda_T^3(T,m)},&T<T_{\rm BEC}\hsp ,
\end{array}\right.
\ee
where $\sigma\approx n/m$ and $z$ is found by solving
the equation \mbox{$n\lambda_T^3(T,m)/g=g_{\,3/2}\hspm (z)$}. The second term in (\ref{cs2b}) is the same as for ideal
Bose-gas\footnote{In the Boltzmann approximation $c_s=\sqrt{5T/(3m)}$.}, and
the first one takes into account
the interaction effects.
%----------------------------------
\begin{figure}[htb!]
\centering
\includegraphics[trim=2cm 8cm 3cm 8cm,width=0.48\textwidth]{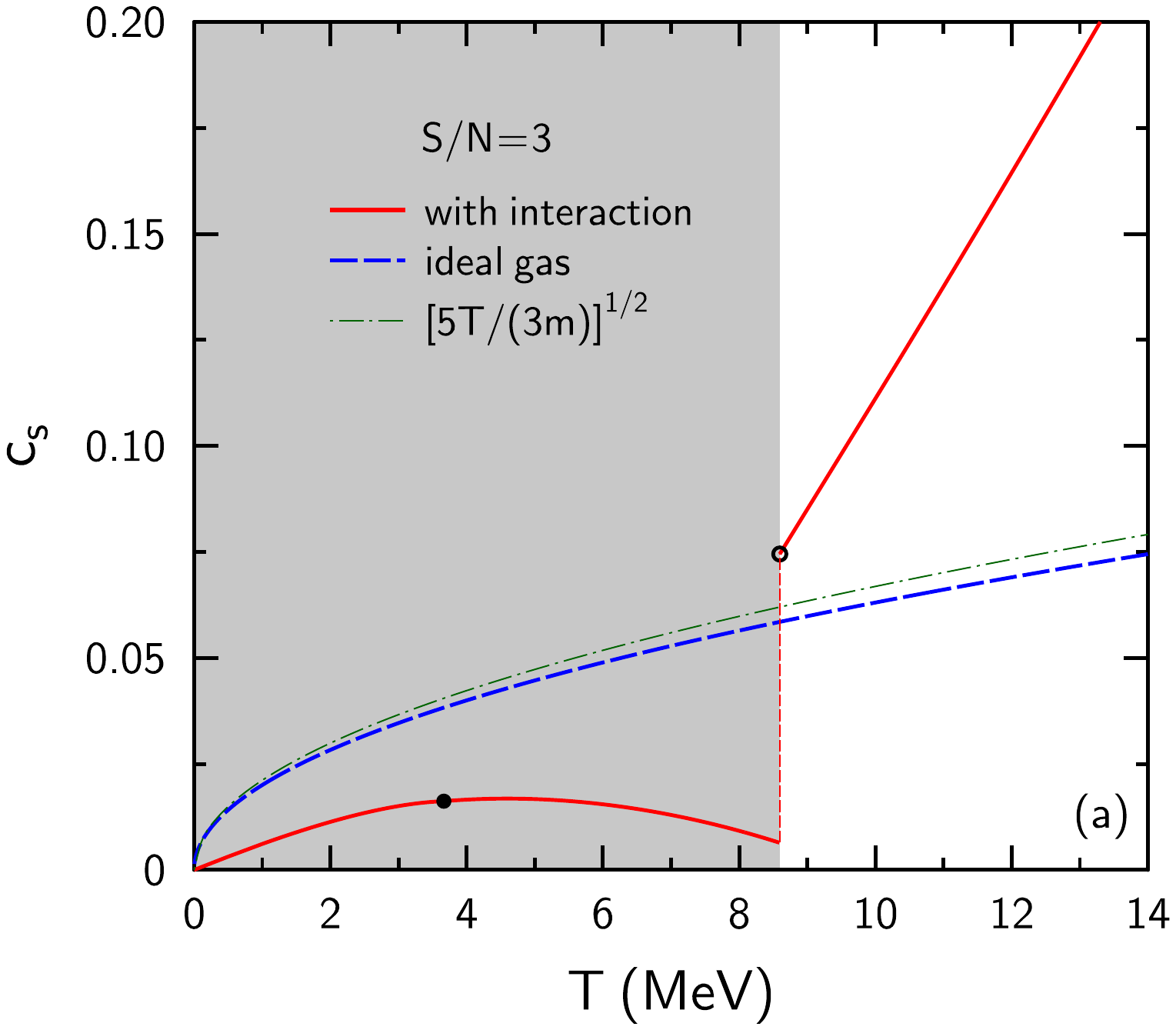}
\includegraphics[trim=2cm 8cm 3cm 8cm,width=0.48\textwidth]{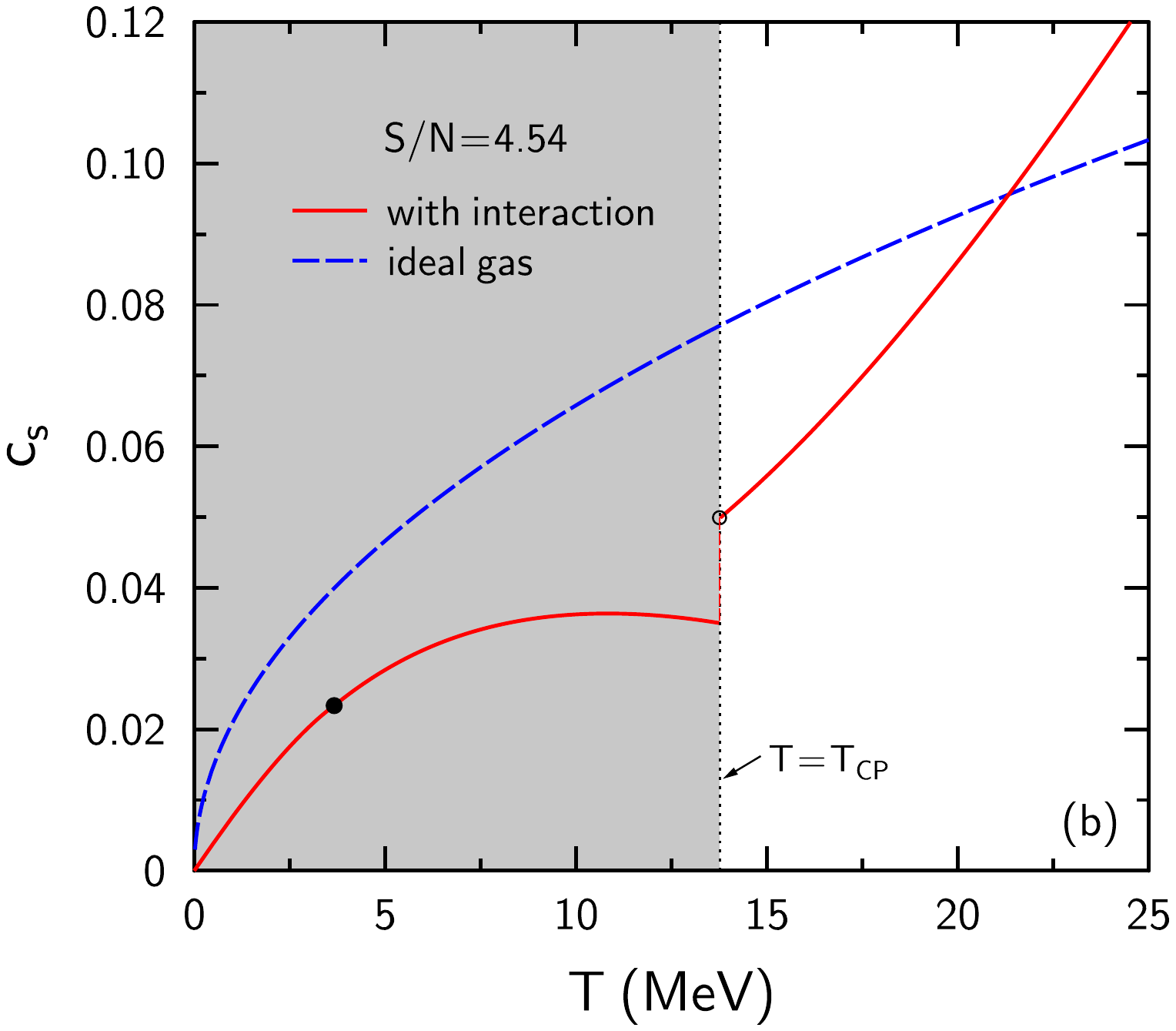}
\caption{
Sound velocity as the function of temperature along the isentropes $S/N=3$ (a)
and $S/N=4.54$ (b)   {(the solid lines). The dashed curves correspond to ideal gas.} The MP states are shown by shading.
{Full dots mark the BEC boundary at $T=T_{\rm TP}$\hspm .}
}\label{fig10}
\end{figure}

The results of calculation are presented in Figs.~\ref{fig10}\hsp (a) and (b) for states along the isentropes  $S/N=3$ and $S/N=4.54$\,,
respectively. As expected, the sound velocity is strongly suppressed inside the MP region.
On the other hand, it becomes much larger at higher densities where the repulsive interaction dominates.
Outside the MP
region $c_s$~increa\-ses with $T$ faster as compared to the ideal-gas calculation. One can see discontinuities
of sound velocities at the MP boundary:
$c_s$ jumps down during  the adiabatic expansion. It is known~\cite{Zel02}
that such a~behavior leads to the formation of a rarefaction shock.
The latter may be regarded as
a~signature of the LGPT.

\subsection{Enhanced heat capacity}
\label{sec-cv}

The isochoric heat capacity is another observable which is sensitive to the LGPT and BEC effects.
This quantity (per particle) is defined as~\cite{Lan75}
\bel{ischc}
c_v=\dfrac{T}{n}\,\left(\dfrac{\partial\hspm s}{\partial T}\right)_n
=\dfrac{1}{n}\,\left(\dfrac{\partial\hspm\varepsilon}{\partial T}\right)_n\,.
\ee
We calculate $c_v$ outside the MP by using formulae similar to
Eqs.~(\ref{ome1}), (\ref{sig4}). For states within the MP we
perform the direct calculation based on Eqs.~(\ref{vfrac}) and (\ref{entd2}).

%------------------
\begin{figure}[htb!]
\centering
\includegraphics[trim=2cm 8cm 3cm 8.5cm,width=0.6\textwidth]{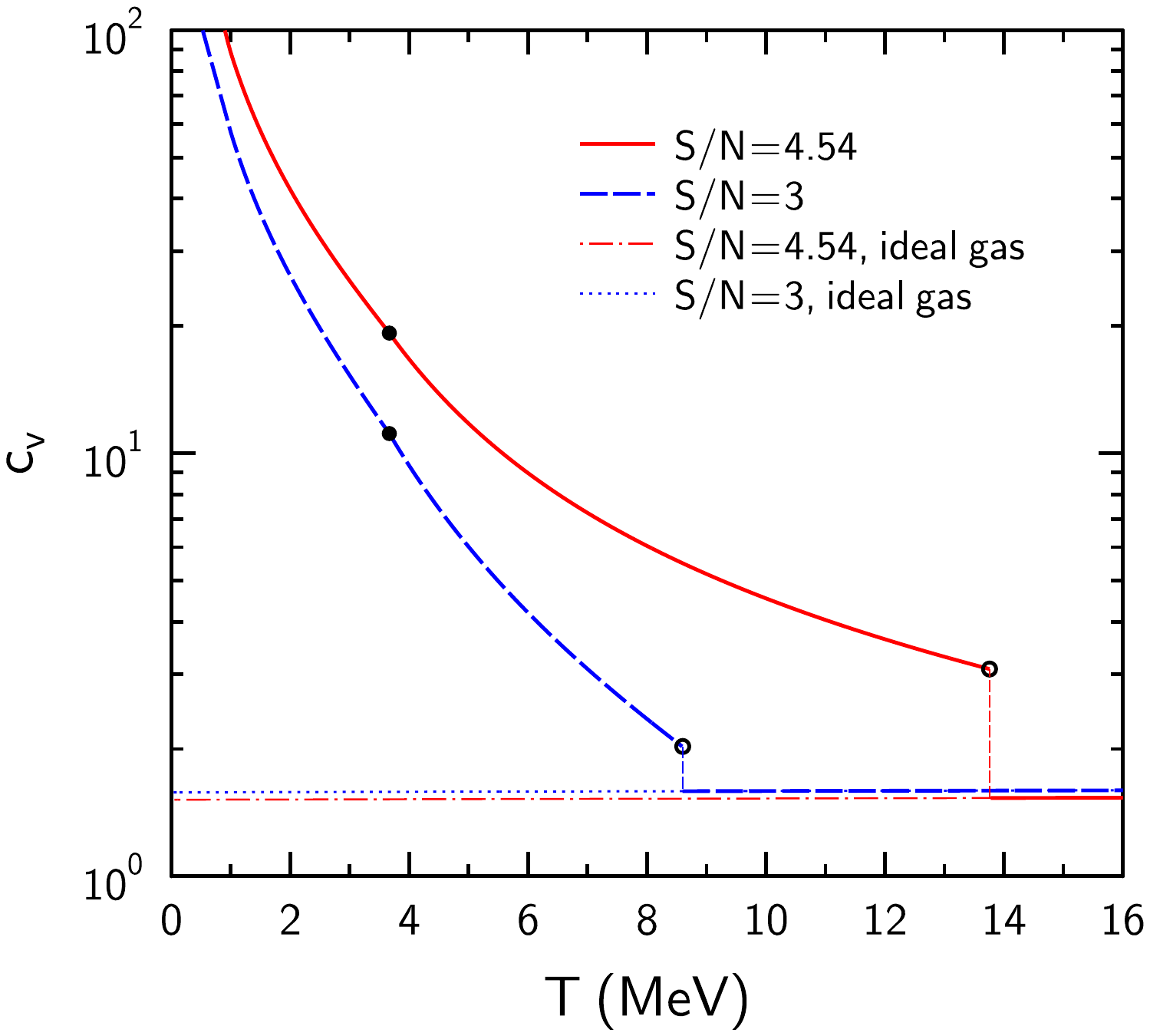}
\caption{
Isochoric heat capacity per particle as the function of temperature along the isentropes \mbox{$S/N=3$}~(dashed)
and \mbox{$S/N=4.54$}~(solid).   {Full dots correspond to $T=T_{\rm TP}$\hspm .
Open dots mark the MP boundary. Thin dotted and dash-dotted lines represent the ideal gas calculation.}
}\label{fig11}
\end{figure}
%------------------
Within the NRA one can use
approximate relations (\ref{snnr}), (\ref{spent}) with \mbox{$M\approx m$}.
Outside the~MP, we obtain the result coinciding with that for the ideal Bose gas~\cite{Pat11}
\bel{ishc1}
c_v\approx\left\{\begin{array}{ll}
\dfrac{15}{4}\hsp\frac{\ds g_{\,5/2}\hspm (z)}{\ds g_{\,3/2}\hspm (z)}-\dfrac{9}{4}\hsp\frac{\ds g_{\,3/2}\hspm (z)}
{\ds g_{\,1/2}\hspm (z)}\,,~~~&T>T_{\rm BEC}\,,\\[3mm]
\dfrac{15}{4}\hspm\frac{\ds\xi\hspm (5/2)}{\ds n\lambda_T^3(T,m)},&T<T_{\rm BEC}\hsp .
\end{array}\right.
\ee
In the Boltzmann approximation one
gets the well-known value \mbox{$c_v=3/2$}.
At the BEC boundary $c_v=3.75\, \xi(5/2)/\hsp\xi(3/2)\approx 1.925$\,.

Figure~\ref{fig11} shows the temperature dependence of $c_v$
along the isentropes with
same $S/N$ as in Fig.~\ref{fig10}. Again one can see jumps of the heat capacity
at the MP boundary. The predicted
$c_v$-values are much larger
in the MP region as compared to the ideal gas calculation. These values increase roughly
as $T^{-1}$ at \mbox{$T\to 0$}.
Such a~behavior can be qualitatively understood.
Indeed, as mentioned above,
at low temperatures
the isentrope
$\widetilde{s}=\textrm{const}$
becomes close to the gas-like
binodal, i.e., $\lambda\approx 1$ and $n\approx n_g(T)$\hsp . Substituting $s\approx\widetilde{s}\hsp n_g\hspm (T)$
into~\re{ischc} gives
\bel{ishc2}
\frac{c_v}{\widetilde{s}}
\approx T\,\frac{n_g^\prime(T)}{n_g(T)}\approx\frac{m-\mu\hspm (T)}{T}
\approx\frac{W_0}{T}\,.
\ee
In the second equality we
neglect deviations from the Boltzmann statistics having  in mind that the density $n_g$
is small at low temperatures.

\section{Conclusions}
  We have proposed a field-theoretical model to describe $\alpha$-clustered nuclear matter
  at finite temperatures. The system of interacting $\alpha$ particles is represented by
a scalar field $\phi$ with the Lagrangian   containing the attractive ($\phi^4$) and repulsive ($\phi^6$)
self-interaction terms. The calculations are done within the mean-field approach which
  obeys a self-consistency relation between the scalar mean
field and the particle effective mass. The model has two   free
parameters   which are fixed by fitting properties of the ground state of cold $\alpha$ matter
known from microscopic calculations of Clark and Wang~\cite{Cla66}.

Our main results are as follows:
1) $\alpha$ matter exhibits a liquid-gas phase transition with
the critical point at $T_c\approx 14~\textrm{MeV},
n_c\approx 0.012~\textrm{fm}^{-3}$;
2)~at low  temperatures the $\alpha$ matter contains
the Bose-Einstein condensate, which appears in the liquid phase; 3) all isentropic trajectories,
$S/N=\textrm{const}$, terminate in the mixed-phase region;
4) the BEC boundary outside the mixed phase
coincides with the isentrope $S/N\approx 1.28$\,;
5) the scaled variance of density fluctuations
has a strong peak at the critical point which lies
on the isentrope $S/N\approx 4.5$\,;
6) the sound velocity and the isochoric heat capacity exhibit jumps at the mixed phase boundary.

 Based on these results, we can formulate a strategy for experimental studies of
\mbox{$\alpha$-clustered} matter. We believe that enhanced yields of $\alpha$ particles and $\alpha$-conjugate nuclei
observed in
Refs.~\cite{Reisdorf:2010aa,Sch17} are associated with the mixed phase of $\alpha$ matter
formed at a late stage of the nuclear matter evolution. Since at this stage the expansion
is approximately isentropic, one may hope to select events which correspond to the critical point of $\alpha$ matter
around $S/N\approx 4.5$. In the vicinity of this
point we expect a wide (power-law) mass distribution of produced fragments,
like $Y(A)\propto A^{-\tau}$~where
$\tau\approx 2$ is the Fischer exponent~\cite{Bondorf:1995ua}\,.

Interesting manifestations of $\alpha$
clustering can be expected in heavy and superheavy nuclei. As demonstrated
in Ref.~\cite{Mis18}, due to strong Coulomb repulsion
such nuclei may develop a hollow structure where
$\alpha$'s are condensed in the outer shell
but neutrons fill the central region.
We hope that future experiments with heavy-ion beams at intermediate energies
will  provide new evidence for $\alpha$ clustering and $\alpha$ condensation in
nuclear systems.

\begin{acknowledgments}
The authors thank D.~Anchishkin, M. Gorenstein, and V. Vovchenko for fruitful
discussions.
L.M.S, R.V.P. and I.N.M.~thank the support from the Frankfurt Institute for Advanced Studies.
H.St.~appreciates the support from J.~M.~Eisenberg Laureatus chair and the W. Greiner Gesellschaft.
\end{acknowledgments}


\begin{thebibliography}{99}

\bibitem{Schuttauf:1996ci}
A.~Schüttauf \textit{et al.},
%``Universality of spectator fragmentation at relativistic bombarding energies,''
Nucl.~Phys. A  \textbf{607}, 457 (1996).

\bibitem{Reisdorf:2010aa}
W.~Reisdorf \textit{et al.} (FOPI Collaboration),
%``Systematics of central heavy ion collisions in the 1A GeV regime,''
Nucl.~Phys. A \textbf{848}, 366 (2010).

\bibitem{Wada:2019utj}
R.~Wada, W.~Lin, P.~Ren, H.~Zheng, X.~Liu, M.~Huang, K.~Yang, and K.~Hagel,\\
%``Experimental liquid-gas phase transition signals and reaction dynamics,''
Phys.~Rev. C \textbf{99}, 024616 (2019).

\bibitem{Gro83}
D.~E.~H.~Gross, Phys. Scripta \textbf{T5}, 213 (1983).

\bibitem{Bon85}
J.~P.~Bondorf, I.~N.~Mishustin, C. J. Pethick, H. Schultz, and K. Sneppen,
Nucl.~Phys.~A~\textbf{443}, 321 (1985).

\bibitem{Peilert:1991sm}
G.~Peilert, J.~Randrup, H.~Stoecker and W.~Greiner,
%``Clustering in nuclear matter at subsaturation densities,''
Phys.~Lett. B \textbf{260}, 271 (1991).

\bibitem{Bondorf:1995ua}
J.~P.~Bondorf, A.~S.~Botvina, A.~S.~Ilinov, I.~N.~Mishustin, and K.~Sneppen,
%``Statistical multifragmentation of nuclei,''
Phys. Rept. \textbf{257}, 133 (1995).

\bibitem{Marini:2015zwa}
P.~Marini \textit{et al.} (INDRA  Collaboration),
%``Signals of Bose Einstein condensation and Fermi quenching in the decay of hot nuclear systems,''
Phys. Lett. B \textbf{756}, 194 (2016).

\bibitem{Borderie:2016tmc}
B.~Borderie et al.
%``Probing clustering in excited alpha-conjugate nuclei,''
Phys. Lett. B \textbf{755}, 475 (2016).

\bibitem{Sch17}
K.~Schmidt~\textit{et al.}, Phys.~Rev.~C~\textbf{95}, 054618 (2017).

\bibitem{Art20}
D.~A.~Artemenkov~\textit{et al.}, Eur.~Phys.~J.~A~\textbf{56}, 250 (2020).

\bibitem{Bar18}
M.~Barbui~\textit{et al.} (FOPI Collaboration), Phys. Rev. C \textbf{98}, 044601 (2018).

\bibitem{Sch16}
P.~Schuck, Y. Funaki, H. Horiuchi, G. R\"opke, A. Tonsaki, and T. Yamada,
Phys. Scripta \textbf{91}, 123001 (2016).

\bibitem{Cao20}
X.~G.~Cao~\textit{et al.}, JPS Conf. Proc. \textbf{32}, 010038 (2020).


%---------------start of new references--------------------------------
\bibitem{Cla66}
J.~W.~Clark and T.-P. Wang, Ann. Phys.~\textbf{40}, 127 (1966).

\bibitem{Joh80}
M.~T.~Johson and J.~W.~Clark, Kinam~\textbf{2}, 3 (1980).

\bibitem{Hor06}
C.~J. Horowitz and A. Schwenk, Nucl.~Phys.~A~\textbf{776}, 55 (2006).

\bibitem{Sed17}
X.-H. Wu, S.-B. Wang, A. Sedrakian, and G. R\"opke, J. Low Temp. Phys.
\textbf{189}, 133 (2017).

\bibitem{Sed20}
A.~Sedrakian, Eur.~Phys. J. A \textbf{56}, 258 (2020).


\bibitem{Fre18}
M.~Freer, H.~Horiuchi, Y. Kanada-En'yo, D. Lee, and U.-G- Mei\ss ner,
Rev.~Mod.~Phys.~\textbf{90}, 035004 (2018).

\bibitem{Hor91}
H. Horiuchi, Nucl.~Phys.~A~\textbf{522}, 257c (1991).

\bibitem{Fel90}
H. Feldmeier, Nucl.~Phys.~A~\textbf{515}, 147 (1990).

\bibitem{Pie92}
S.~C.~Pieper, R.~B.~Waringa, and V.~R.~Pandharipande,
Phys. Rev. C {\bf 46}, 1741 (1992).

\bibitem{Epe11}
E. Epelbaum, H. Krebs, D. Lee, and U.-G- Mei\ss ner,
Phys.~Rev.~Lett.~\textbf{106}, 192501 (2011).

\bibitem{Elh16}
S. Elhatisari \textit{et al.}, Phys.~Rev.~Lett.~\textbf{117}, 132501 (2016).

\bibitem{Ebr12}
J.-P. Ebran, E. Khan, T. Niksic, and D. Wretenar, Nature~\textbf{487},
341 (2012).

\bibitem{Ebr14}
J.-P. Ebran, E. Khan, T. Niksic, and D. Wretenar,
Phys. Rev. C {\bf 90}, 054329 (2014).

\bibitem{Ebr20}
J.-P. Ebran, M.~Girod, E. Khan, R.~D.~ Losseri, and P.~Schuck,
Phys. Rev. C {\bf 102}, 014305 (2020).

%----------------end of new references------------------------------

\bibitem{Mis16}
S.~Misicu, I.~N.~Mishustin, and W.~Greiner,
Mod.~Phys.~Lett. A~\textbf{32},  1750010 (2016).

\bibitem{Pai18}
H.~Pais, F.~Gulminelli, C.~Provid\'encia, and G.~R\"opke,
Phys. Rev. C {\bf 97}, 045805 (2018).

\bibitem{Zha19}
Z.-W.~Zhang and L.~W.~Chen, Phys. Rev. C {\bf 100}, 054304 (2019).

\bibitem{Fur20}
S.~Furusawa and  I.~N.~Mishustin, Nucl.~Phys.~A~\textbf{1002}, 121991 (2020).


\bibitem{Pei92}
G.~Peilert, J.~Konopka, H.~St\"ocker, W.~Greiner. M.~Blann,
and M.~G.~Mustafa, Phys. Rev. C~\textbf{46}, 1457 (1992).

\bibitem{Mar98}
T. Maryama~\textit{et al.}, Phys. Rev. C~\textbf{57}, 655 (1998).

\bibitem{Sat17}
L.~M.~Satarov, M.~I.~Gorenstein, A.~Motornenko, V.~Vovchenko, I.~N.~Mishustin, and H.~Stoecker,
%``Bose-Einstein condensation and liquid-gas phase transition in alpha-matter,''
J. Phys. G: Nucl. Part. Phys. \textbf{44}, 125102 (2017).

\bibitem{Sat19}
L. M. Satarov, I. N. Mishustin, A. Motornenko, V. Vovchenko, M. I. Gorenstein,\\
and H.~Stoecker, Phys. Rev. C {\bf 99}, 024909 (2019).

\bibitem{Sat20}
L. M. Satarov, M. I. Gorenstein, I. N. Mishustin,  and H.~Stoecker,
Phys. Rev. C {\bf 101},\\ 024913 (2020).

\bibitem{Bog77}
J. Boguta and A. R. Bodmer, Nucl. Phys. A~\textbf{292}, 413 (1977).

\bibitem{Bog83}
J. Boguta and H. Stoecker, Phys. Lett. \textbf{B 120}, 289 (1983).

\bibitem{Mis19}
I.~N.~Mishustin, D.~V.~Anchishkin, L.~M.~Satarov, O.~S.~Stashko, and H.~Stoecker,
%``Condensation of interacting scalar bosons at finite temperatures,''
Phys. Rev. C \textbf{100}, 022201 (2019).


\bibitem{Bay01}
G.~Baym, J.-P. Blaizot, M. Holzmann, F. Lalo\"e, and D. Vautherin,
Eur. Phys.~J.~B~\textbf{24},\\ 107 (2001).

\bibitem{Pat11}
R. K. Pathria and P. D. Beale, \textit{Statistical Mechanics}, (Elsevier, Amsterdam, 2011).

\bibitem{Lan75}
L.~D. Landau and E.~M. Lifshitz, \textit{Statistical Physics} (Pergamon, Oxford, 1975).

%\cite{Savchuk:2020yxc}
\bibitem{Sav20}
O.~Savchuk, Y.~Bondar, O.~Stashko, R.~V.~Poberezhnyuk, V.~Vovchenko, M.~I.~Gorenstein, and H.~Stoecker,
Phys. Rev. C \textbf{102}, 035202 (2020).

%\bibitem{Cla66}
%J.~W.~Clark and T.-P. Wang, Ann. Phys.~\textbf{40}, 127 (1966).

\bibitem{Lan53}
L.~D. Landau, Izv. Akad. Nauk SSSR, Ser. Fiz. \textbf{17}, 51
(1953); in \textit{Collected Papers of L.~D.~Landau}
(Gordon and Breach, New York, 1965), p.~569.

\bibitem{Cle98}
J. Cleymans and K. Redlich, Phys. Rev. Lett. \textbf{81},
5284 (1998).

\bibitem{And06}
A. Andronic, P. Braun--Munzinger, and J. Stachel,
Nucl. Phys. A~\textbf{772}, 167 (2006).

\bibitem{Buc90}
H.~Buchenau, E.~L.~Knuth, J.~Northby, J.~P. Toennies, and C.~Winkler,
J.~Chem.~Phys.~\textbf{92}, 6875 (1990).

\bibitem{Vov15}
V.~Vovchenko, D.~V.~Anchishkin, M.~I.~Gorenstein, J. Phys. A: Math. Theor. \textbf{48}, 305001 (2015).

\bibitem{Pob20}
R.~V.~Poberezhnyuk, O.~Savchuk, M.~I.~Gorenstein, V.~Vovchenko, and H.~Stoecker,
arXiv:~2011.06420 [hep-ph].

\bibitem{Zel02}
Ya. B. Zel'dovich and Yu. P. Reizer,
\textit{Physics of Shock Waves and High-Temperature Hydrodynamic
Phenomena} (Academic Press, New York, 1966/1967).

\bibitem{Mis18}
S.~Misicu, I.~N.~Mishustin, in
\textit{Walter Greiner Memorial Volume}, eds. O.~Hess,
H.~St\"ocker, (World Scientific, Singapore, 2018), p.~263;
arXiv: 1806.01886 \mbox{[nucl-th]}.



\end{thebibliography}
\end{document}